\documentclass[english,aps,twocolumn]{revtex4-1}
\usepackage[T1]{fontenc}
\usepackage[latin9]{inputenc}
\usepackage{color}
\usepackage{babel}
\usepackage{textcomp}
\usepackage{amsmath}
\usepackage{amssymb}
\usepackage{graphicx}
\usepackage[unicode=true,pdfusetitle,bookmarks=true,bookmarksnumbered=false,bookmarksopen=false,breaklinks=false,pdfborder={0 0 0},pdfborderstyle={},backref=false,colorlinks=true]{hyperref}
\hypersetup{linkcolor=blue, citecolor=blue}
\begin{document}
\title{Bursting and excitability in neuromorphic resonant tunneling diodes}
\author{Ignacio Ortega$^{1,2}$}
\author{Oreste Piro$^{1}$}
\author{Bruno Romeira$^{3}$}
\author{Julien Javaloyes$^{1,2}$}
\email{julien.javaloyes@uib.es}

\affiliation{$^{1}$ Departament de Física, Universitat de les Illes Balears, Cra.\,\,de
Valldemossa, km 7.5, E-07122 Palma de Mallorca, Spain}
\affiliation{$2$ Institute of Applied Computing and Community Code (IAC-3), Cra.\,\,de
Valldemossa, km 7.5, E-07122 Palma de Mallorca, Spain}
\affiliation{$^{3}$Ultrafast, Bio- and Nanophotonics, INL - International Iberian
Nanotechnology Laboratory, Av. Mestre José Veiga s/n, 4715-330 Braga,
Portugal}
\date{\today}
\begin{abstract}
We study in this paper the dynamics of quantum nanoelectronic resonant
tunneling diodes (RTDs) as excitable neuromorphic spike generators.
We disclose the mechanisms by which the RTD creates excitable all-or-nothing
spikes and we identify a regime of bursting in which the RTD emits
a random number of closely packed spikes. The control of the latter
is paramount for applications in event-activated neuromorphic sensing
and computing. Finally, we discuss a regime of multi-stability in
which the RTD behaves as a memory. Our results can be extended to
other devices exhibiting negative differential conductance.
\end{abstract}
\maketitle
Spike information processing and transmission in the form of events
that occur at continuous times has numerous advantages over digital
encoding and signaling. It is a key mechanism in the dynamics of neurons
and the brain, which suggests its value in the development of biologically-inspired
artificial intelligence. Neurons are \emph{excitable systems}; they
respond to an external stimulus by realizing a large amplitude response,
typically in the millisecond and millivolt range, before returning
to their rest state, provided that said stimulus is larger than a
certain threshold. If this input is not sufficiently strong, a weak,
exponentially decaying, response is obtained. For the duration of
the response --known as lethargic time-- the system is unable to
respond to any other stimulus, irrespective of its amplitude \citep{Izhikevich,HH-JOP-52,HH2-JOP-52}.
The concept of excitability pervades many areas such as image processing
\citep{kuhnert89}, semiconductor structures \citep{SNJ-JAP-11} and
lasers \citep{GHR-PRL-07,SBB-PRL-14,BKY-OL-11}.

Several neuromorphic circuits have been proposed which attempt to
emulate the transmission of information in the brain and the nervous
systems, including the IBM TrueNorth chip \citep{MAA-SCI-14} and
the Intel Quark SE chip \citep{Intel-SEEIM-16}. These approaches
are still based on adapting the conventional CMOS architecture, and
have some drawbacks, such as low frequency (kHz) and much higher power
consumption than the brain. For low energy, sub-pJ/spike, synaptic-like
functionalities, non-volatile materials such as resistive random-access
memory, phase-change memory and spin-transfer torque magnetic random-access-memory
offer alternatives to silicon, see \citep{ZZYH-APR-20} for a review.
Despite remarkable progresses, fan-out and parasitic constraints of
these approaches limit the power budget and frequency operation for
scalable, high-speed solutions. 

In this work, we provide a detailed analysis on the performance of
quantum nanoelectronic resonant tunneling diodes (RTDs) as neuromorphic
spike generators. Resonant tunneling diodes are promising candidates
and are the fastest electronic oscillators up to date, reaching frequencies
in the order of the hundreds of GHz, with a world record of 1.98 THz
\citep{ISA-IEEE-17,RFJ-CHAOS-18}. Their speed stems from the nanometric
size ($\sim10\,$nm) of the semiconductor active layer of the RTD
in the epitaxial growth direction. This active layer consists of a
double barrier quantum well (DBQW) nanostructure. This provides to
RTD devices a current voltage with pronounced negative differential
conductance which has been extensively applied for oscillator devices.
Wang et al reported RTD oscillators with areas of $15$ and $25\,\mu$m$^{2}$
operating at powers under 1 mW \citep{WAO-IEEE-15}. Asada's group
reports a transmitter comprising a $1\,\mu$m$^{2}$ RTD with a maximum
oputput power of $60\,\mu$W \citep{OHH-IEEE-16,OHS-IEEE-16}. RTDs
prospective applications as both transmitters and receivers of digital
coding are being extensively investigated, as they can be modulated
via either amplitude shift keying (ASK) or on-off keying (OOK). Recently,
short-distance wireless data transmission at about 10 Gbps using RTD-based
devices has been achieved \citep{DNN-IEEE-17,WAW-IEEE-18}. For neuron
computation, early works evoked RTD-based devices as potential nanoelectronic
candidates for cellular neural networks as a form of threshold logical
gates \citep{HC-CTA-01}.

The Hodgkin-Huxley, integrate and fire, or the Izhikevich models are
widely used as test benches for neural spiking. They provide for different
compromises between biological accuracy and computational cost. In
comparison, Liénard oscillators \citep{RJF-IEEE-13} such as RTDs
received comparatively less attention. The excitability of RTDs was
disclosed in \citep{RJI-OE-13} both using electrical and optical
stimuli, and optical spike regeneration coupling the RTD to laser
diode and a time delayed feedback loop was demonstrated in \citep{RAF-SR-16}.
Yet, a full theoretical characterization of the spiking and bursting
dynamics of RTD devices is lacking. Importantly, the results presented
here can be applied to other negative devices exhibiting differential
conductance. Some recent examples include NDC devices using Van der
Walls \citep{LGA-Nature-15} or graphene/boron nitride \citep{MTC-Nature-14}
heterostructures.

\begin{figure}[t]
\centering{}\includegraphics[width=1\columnwidth]{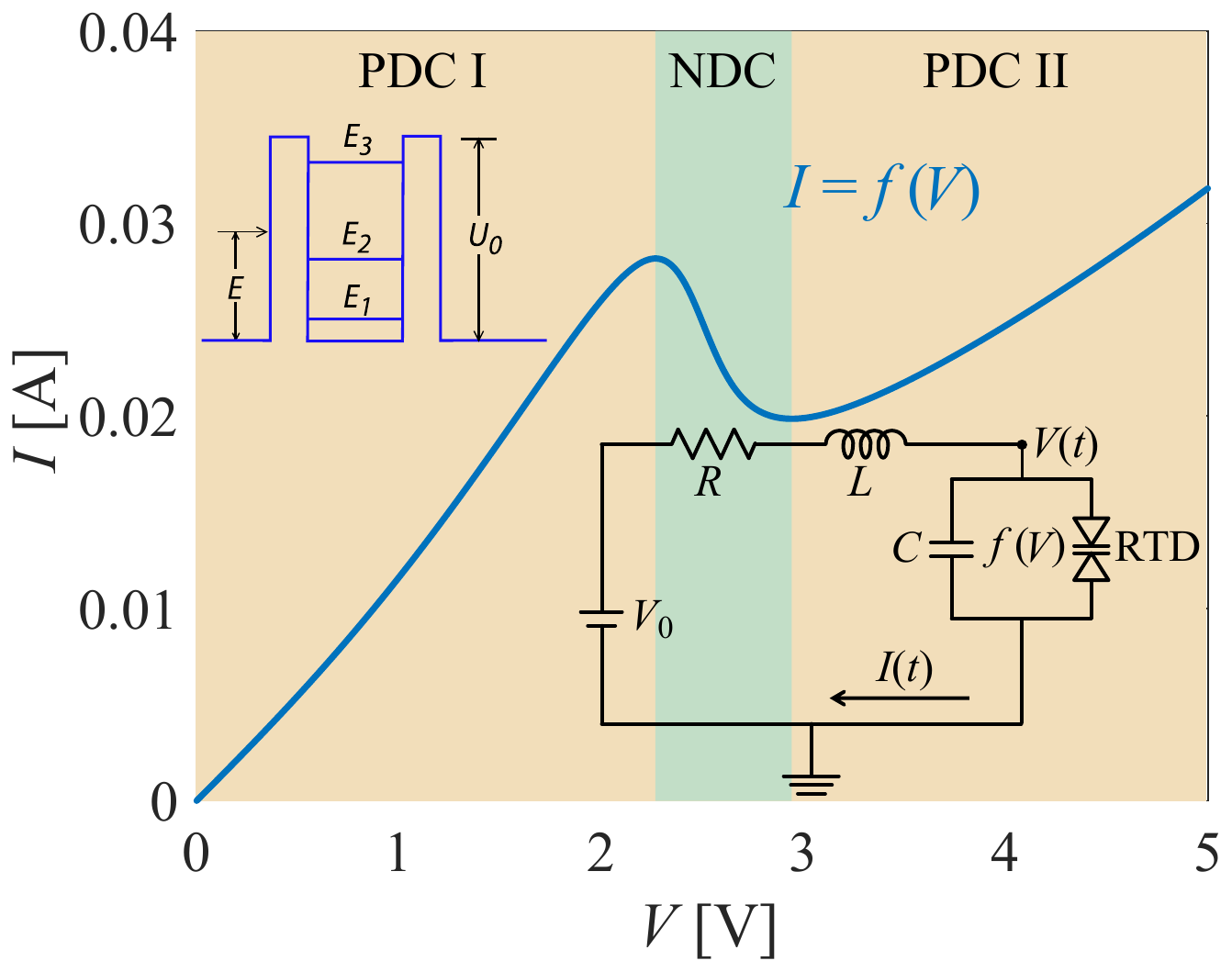}\caption{Current-voltage nonlinear relationship of the double barrier quantum
well resonant tunneling diode. The regions with positive differential
conductance (PDC I, PDC II) and negative differential conductance
(NDC) are also illustrated. Upper inset: A symmetric double barrier
quantum well nanostructure where $U_{o}$ is the potential barrier
height, and $E_{1}$ - $E_{2}$ are the resonant energy levels. Lower
inset: Schematic circuit of the RTD oscillator.\label{fig:I-V_curve}}
\end{figure}

The paper is organized as follows: In section \ref{sec:theoretical_model}
we recall the fundamentals of nanoelectronic RTD devices and the basic
hypothesis underlying our modeling approach and the various types
of solutions. In section \ref{sec:RTD_excitable} we give a detailed
analysis of the excitable response of the RTD and discuss slow-fast
dynamics and bursting. Finally, section \ref{sec:bifurcation} is
devoted to the bifurcation analysis of our dynamical system, i.e.,
the qualitative changes of the system's responses as a function of
its control parameters. We summarize our results in section \ref{sec:conclusions}.

\section{Theoretical model \label{sec:theoretical_model}}

The active layer of nanoelectric RTD devices consists of a narrow
bandgap semiconductor embedded between two thin layers of wide bandgap
semiconductors, thus forming a double barrier quantum well (DBQW)
nanostructure (figure \ref{fig:I-V_curve}, upper inset). Depending
on the voltage applied across the device, the Fermi level of the incident
electrons may resonate with the confinement levels of the quantum
well, thus locally maximizing their probability to cross it (i.e.,
maximum probability corresponds to a local peak of current). As a
result, the current-voltage characteristic of the RTD is a nonlinear
function with one or more regions of negative differential conductance
in between regions of positive differential conductance \citep{WAW-IEEE-18},

\begin{eqnarray}
f\left(V\right) & = & a\ln\left(\frac{1+e^{\frac{q}{k_{B}T}\left(b-c+n_{1}V\right)}}{1+e^{\frac{q}{k_{B}T}\left(b-c-n_{1}V\right)}}\right)\label{eq:I-V_curve}\\
 & \times & \left(\frac{\pi}{2}+\tan^{-1}\left(\frac{c-n_{1}V}{d}\right)\right)+h\left(e^{\frac{q}{k_{B}T}n_{2}V}-1\right).\nonumber 
\end{eqnarray}

The expression \ref{eq:I-V_curve} is derived by applying the Fermi-Dirac
statistics to a double barrier quantum well. $T$ is the temperature,
$q$ is the electron charge and $k_{B}$ is the Boltzmann's constant.
The inner parameters $a$, $b$, $c$, $d$, $n_{1}$, $n_{2}$ and
$h$ depend on the geometry of the barrier and its resonant energy
levels, although they can also be determined by fitting experimental
data \citep{SDC-EDL-96}. In this article, the inner parameters have
the following values: $a$ = 0.0039 A, $b$ = 0.05 V, $c$ = 0.0874
V, $d$ = 0.0073 V, $n_{1}$ = 0.0352, $n_{2}$ = 0.0031, $h$ = 0.0367
A, $T$ = 300°K. These parameters produce an I-V characteristic typical
of III-V semiconductor RTD devices (e.g. InGaAs/ALAs materials) and
with micrometric overall device size. A plot of the function $f(V)$
with these values is shown in figure \ref{fig:I-V_curve}. This characteristic
is referred to as N-shaped since it has a region of negative slope
(or conductance) embedded between two regions of positive conductance.
We will refer to these regions as NDC, PDC I and PDC II.

A system of two first-order differential equations for the current
and voltage was proposed in \citep{RJF-IEEE-13}, accounting for the
dynamics of a double barrier quantum well resonant tunneling diode
(DBQW RTD) connected to a DC voltage input. A schematics of the circuit
is shown in figure \ref{fig:I-V_curve}, lower inset. The dynamics
equations are derived from Kirchoff laws,
\begin{eqnarray}
\mu\dot{V} & = & I-f(V),\label{eq:dVdt}\\
\mu^{-1}\dot{I} & = & V_{0}-V-RI.\label{eq:dIdt}
\end{eqnarray}

Here, $V(t)$ is the voltage across the RTD and $I(t)$ is the total
current. $V_{0}$ is the bias DC voltage, $R$ is the circuit intrinsic
resistance and the parameter $\mu$ is defined as $\mu=\sqrt{\frac{C}{L}}$,
where $L$ and $C$ are the equivalent inductance and capacitance,
respectively, which sets the circuit's natural frequency. This parameter
reduction has been achieved by defining a dimensionless time $t=\omega_{0}\tilde{t}$,
setting $\dot{X}\equiv dX/dt$ and where $\omega_{0}=\frac{1}{\sqrt{LC}}$
is the RTD natural frequency. We note that Eqs.~(\ref{eq:dVdt},\ref{eq:dIdt})
represents a Liénard oscillator \citep{Strogatz,L-RGE-28,RJF-IEEE-13}.

\begin{figure}[t]
\centering{}\includegraphics[width=1\columnwidth]{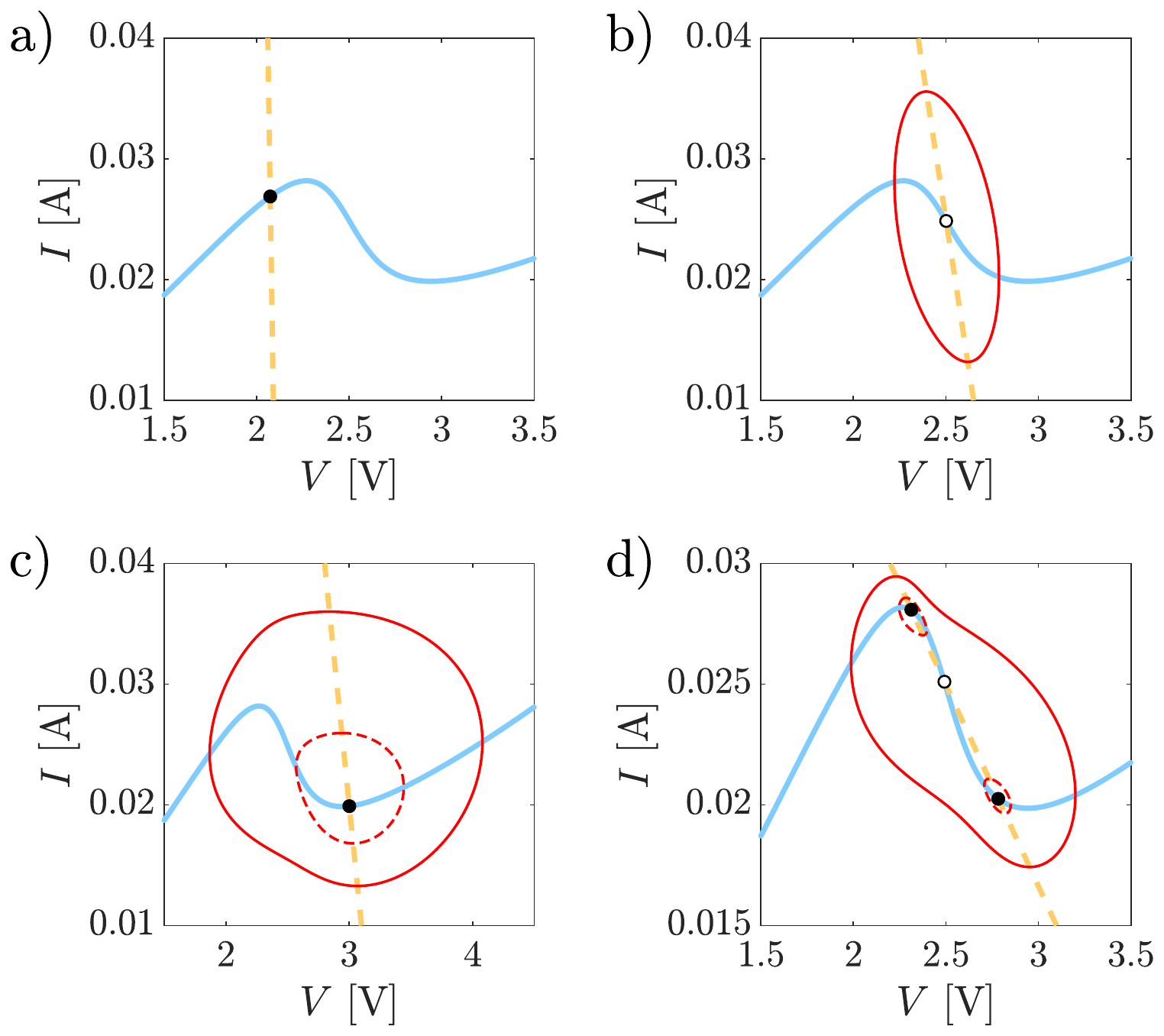}\caption{Stable and unstable solutions in equilibrium of equations (\ref{eq:dVdt},\ref{eq:dIdt})
under different parameters. The blue and yellow lines correspond to
the conditions $\dot{V}=0$ and $\dot{I}=0$, respectively. The intersections
correspond to fixed points. Stable (resp. unstable) periodic orbits
are depicted in continuous (resp. dotted) red lines: a) $R=1\,\Omega$,
$\mu=0.04\,\Omega^{-1}$, $V_{0}=2.1\,\text{V}$.\emph{ b}) $R=10\,\Omega$,
$\mu=0.04\,\Omega^{-1}$, $V_{0}=2.75\,\text{V}$.\emph{ c}) $R=10\,\Omega$,
$\mu=0.01\,\Omega^{-1}$, $V_{0}=3.2\,\text{V}$. \emph{d}) $R=60\,\Omega$,
$\mu=0.01\,\Omega^{-1}$, $V_{0}=3.998\,\text{V}.$\label{fig:limit_cycles}}
\end{figure}

After proper normalization, there are only three parameters, which
depend on the circuit design: the resistance $R$, the stiffness parameter
$\mu$ and the bias $V_{0}$. As discussed in section \ref{subsec:slowfast},
sufficiently small values of $\mu$ are necessary for the RTD to exhibit
excitability. Depending on the parameters $R$, $\mu$ and $V_{0}$,
the system given by Eqs.~(\ref{eq:dVdt},\ref{eq:dIdt}) may exhibit
fixed points and periodic solutions, as well as a coexistence between
two or more equilibrium solutions. Some examples of coexisting solutions
for different choices of parameters are illustrated in figure \ref{fig:limit_cycles}.

\textbf{Fixed point:} The system exhibits an equilibrium point wherever
the nullclines intersect (figure \ref{fig:limit_cycles}). In other
words, the following equations must be satisfied:
\begin{eqnarray}
f(V)-I & = & 0,\label{eq:I-V_nullcline}\\
V+RI-V_{0} & = & 0,\label{eq:load_line}
\end{eqnarray}

which implies solving the nonlinear equation $V+Rf(V)=V_{0}$. The
stability of a fixed point is given by the eigenvalues of the Jacobian
of (\ref{eq:dVdt},\ref{eq:dIdt}) that reads
\begin{equation}
J=\begin{bmatrix}-\frac{1}{\mu}f'(V) & \frac{1}{\mu}\\
-\mu & -\mu R
\end{bmatrix}.\label{eq:Jacobian}
\end{equation}

We have 
\begin{eqnarray}
\lambda_{\pm} & = & -\frac{1}{2}\left(\tfrac{f'(V)}{\mu}+\mu R\right)\pm\frac{1}{2}\sqrt{\left(\tfrac{f'(V)}{\mu}-\mu R\right)^{2}-4}.\label{eq:eigenvalues}
\end{eqnarray}

The fixed point is stable if the trace is negative and the determinant,
positive. For this system, the aforementioned inequalities read,
\begin{eqnarray}
f'(V) & > & -\mu^{2}R,\label{ineq:trace}\\
f'(V) & > & -\frac{1}{R}.\label{ineq:determinant}
\end{eqnarray}

When they are saturated, these two inequalities correspond to the
locus bistability via Saddle-Node bifurcations and to the creating
of limit cycles via Andronov-Hopf bifurcations, respectively. It is
clear from these inequalities that being either in the PDC I or the
PDC II region is a sufficient (although not necessary) condition for
a fixed point to be stable since $f'(V)>0$. The number of steady
states is determined to a great extent by whether the resistance $R$
is larger than a critical value, given by the absolute value of the
reciprocal of the minimal conductance,
\begin{equation}
R_{\text{\text{C}}}=-\frac{1}{\min\left\{ f'(V):V\in\mathbb{R}\right\} }.\label{eq:R_crit}
\end{equation}

If $R\leq R_{C}$, the system has a unique fixed point for all values
of $\mu$ and $V_{0}$. If $R>R_{C}$ the system has between one and
three fixed points, as the load line may intersect the I-V characteristic
in up to three points. This is geometrically intuitive but it will
also be discussed in part \ref{sec:bifurcation}. For the typical
parameter set used in our analysis, the critical resistance is $R_{C}\simeq38.484\,\Omega$.

\textbf{Periodic solution:} The system (\ref{eq:dVdt},\ref{eq:dIdt})
may also exhibit one or more periodic solutions which may be stable
or unstable. The stability of a periodic solution is given by the
Floquet multipliers \citep{Strogatz}. A limit cycle may arise from
a stable focus as the latter becomes stable, in an supercritical Andronov-Hopf
(AH) bifurcation \citep{Strogatz,Izhikevich}. Figure \ref{fig:limit_cycles}\emph{b
}Shows a stable limit cycle surrounding an unstable fixed point. Simulations
of equations (\ref{eq:dVdt},\ref{eq:dIdt}) also show a stable limit
cycle in coexistence with a stable fixed point as well as an unstable
limit cycle in between (figure \ref{fig:limit_cycles}\emph{c}). This
suggests the existence of a subcritical AH bifurcation. As mention
above, a circuit with high resistance ($R>R_{C}$) may exhibit multiple
fixed points and, with them, multiple limit cycles. A particular case
of this is depicted in figure \ref{fig:limit_cycles}\emph{d.}

\section{excitability in an RTD circuit\label{sec:RTD_excitable}}

\subsection{Slow-fast dynamics\label{subsec:slowfast}}

\begin{figure}[t]
\centering{}\includegraphics[width=1\columnwidth]{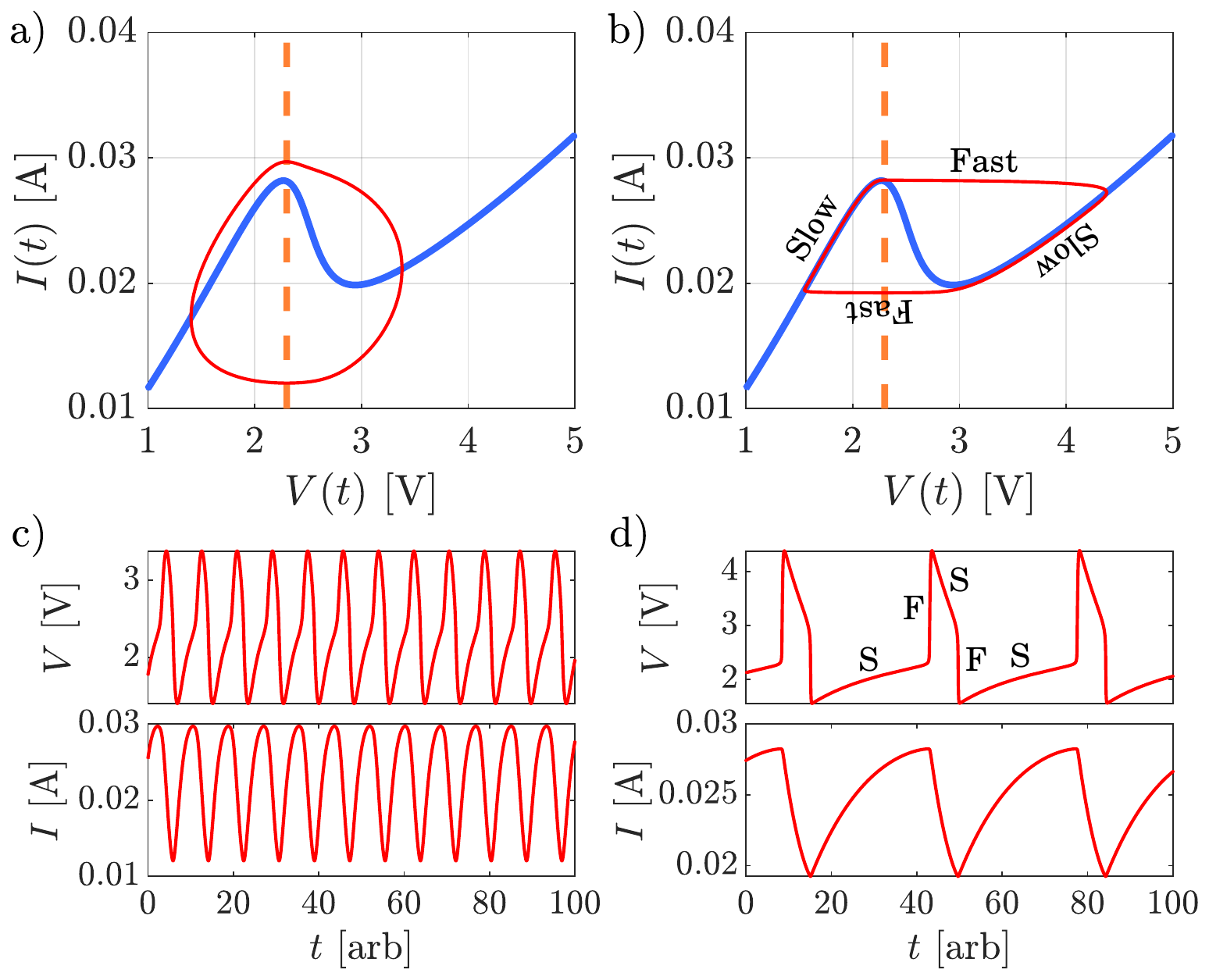}\caption{Periodic solution on the phase space for $R=0.01\,\Omega$, $V_{0}=2.3\,\text{V}$,
and two different values for the stiffness coefficient: \emph{a})
$\mu=0.008\,\Omega^{-1}$, \emph{b}) $\mu=0.001\,\Omega^{-1}$. The
I-V curve (blue line) and load line (dashed orange line) are included.
\emph{b,c}) Evolution of over time of the output variables $V$ and
$I$ corresponding to the periodic solutions in panels \emph{a} and
\emph{b}, respectively. The system exhibits stages of slow and fast
dynamics when $\mu$ is small enough. \label{fig:slowfast}}
\end{figure}

As mentioned in part \ref{sec:theoretical_model}, the parameter $\mu$
is referred to as the stiffness coefficient and it is of key importance
because it determines whether or not the circuit is suitable for spike
generation. If $\mu\sim1$, the periodic solutions are well--rounded
orbits, with the coordinates $V(t),I(t)$ evolving over time in quite
smooth fashion, as shown in the example in figure \ref{fig:slowfast}\emph{a,c}
and Fig.~\ref{fig:limit_cycles}. However, as $\mu$ is decreased,
the orbits become stiffer. If $\text{\ensuremath{\mu}}$ is sufficiently
small (several times smaller than $1/R_{C}$), four stages can be
recognized in each period; two slow stages and two fast stages, as
it is the case with the periodic solution shown in figure \ref{fig:slowfast}\emph{b,d}.
In the first slow stage, the orbit overlaps the N-shaped RTD current-voltage
characteristic in the PDC I, and $V(t)$ and $I(t)$ increase steadily.
This means that $I\approx f(V)$, i.e., all the incident electrons
are crossing the double barrier quantum well. When the local maximum
of the I-V characteristic is reached, the first fast stage initiates,
and the voltage suddenly rises. This happens because far from the
I-V curve, if $\mu$ is very small then any change in $V$ is significantly
larger than the corresponding change in $I$, which remains constant
and larger than $f(V)$. This means that not all incident electrons
are crossing the barrier, and charges accumulate at the ends. The
orbit reaches the I-V curve at the PDC II in a very short time, and
the second slow stage starts; $V(t)$ and $I(t)$ decrease steadily
as they follow the $\dot{V}=0$ nullcline until reaching its local
minimal point. The second fast stage begins with a sudden voltage
decrease and almost no change in current. Since $I<f(V)$ in this
stage, the ends of the DBQW is discharged. The ends of the DBQW being
charged and discharged are schematically represented in figure \ref{fig:I-V_curve}
(inset) by the intrinsic capacitance \emph{$C$}. The slow and fast
stages can be easily distinguished in each period of the profile of
$V(t)$ and they give it the characteristic of a periodic spike signal.
The variable $I(t)$ on the other hand, does not have a sudden rise
or drop, thus being an overall slow variable.

Because of the two stages of slow dynamics the period of the limit
cycle increases as the stiffness coefficient $\mu$ decreases. Once
time normalization is removed one can appreciate that the period is
proportional to $L$, see Eq.~(6) in \citep{RJI-OE-13}. The exact
limit cycle period depends on the parasitic capacitance and inductance
the device is fabricated with. As a reference, Wang et al report the
fabrication of an RTD with a DBQW of about $7\,$nm width, a mesa
area of $16\,\mu$m$^{2}$ and a parasitic capacitance as low as about
$C=78\,\text{fF}$ \citep{WAW-IEEE-18}. Together with a transmission
line inductance of $L\sim100\,\text{nH},$a stiffness coefficient
of $\mu=\sqrt{\tfrac{C}{L}}\approx0.001\,\Omega^{-1}$ can be achieved.
Under this setting, a normalized refractory time in the order of $T\sim10$
like in Fig.~(\ref{fig:slowfast}\emph{b)} would translate into a
frequency in the order of the GHz. This value is two orders of magnitude
lower than the frequencies typically reported for RTDs with similar
dimensions \citep{WAO-IEEE-15,DNN-IEEE-17,ISA-IEEE-17}. It is because
these devices are fabricated to operate as oscillators in smooth dynamics
with higher values of $\mu$.

\subsection{Excitable response\label{subsec:excitable}}

\begin{figure}[t]
\centering{}\includegraphics[width=1\columnwidth]{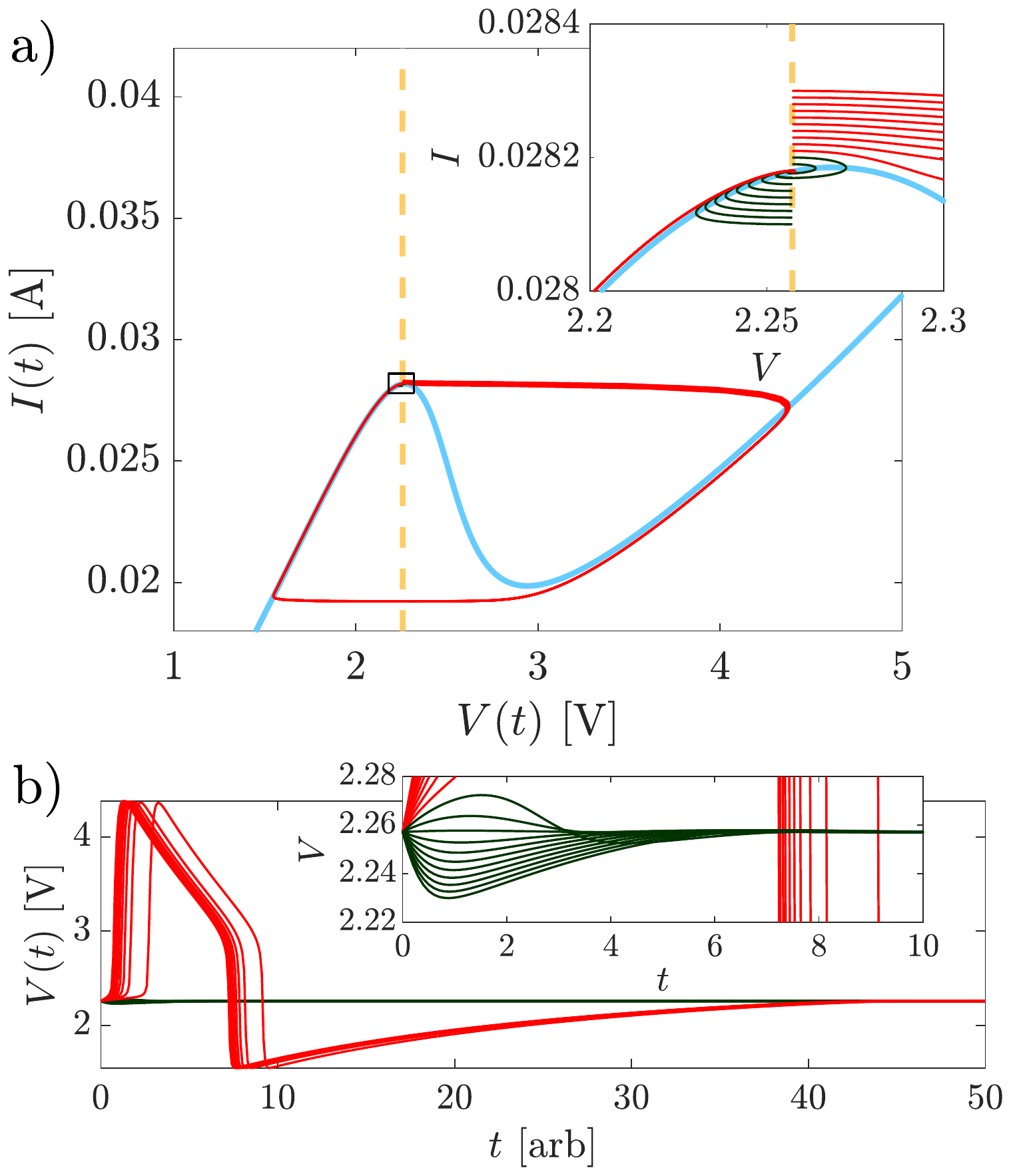}\caption{System\textquoteright s response to several perturbations out of its
natural equilibrium for parameters $R=0.1\,\Omega$, $\mu=0.001\,\Omega^{-1}$,
$V_{0}=2.26\,\text{V}$. \emph{a}) Trajectories on the phase plane
$(V,I)$. \emph{b}) Evolution of variables $V$ and $I$ over time.
Current perturbations under 0.0282 A induce an exponential decay (green
curves and insets). Perturbations above this threshold lead to the
generation of a single spike before returning to the attractor (red
curves).\label{fig:excitable}}
\end{figure}

Let us consider a configuration in the circuit where the load line
intersects the RTD I-V curve in a unique point either in the PDC I
or II but close to the NDC (i.e., close to either the local maximum
or minimum). In this situation, the intersection point is a stable
attractor. Figure \ref{fig:excitable} shows several responses of
the system after perturbations from its stable point of equilibrium.
Indeed, if the perturbation is above a certain threshold ($\Delta I\simeq0.0282\,$A
here), the system exhibits a single orbit, thus producing a spike.
The orbit is reminiscent of the stable limit cycle obtained for slightly
different values of the bias when the RTD is biased in the NDR. On
the other hand, the response to a weak perturbation is a small response
that decays exponentially.

\subsection{Spikes and bursts\label{subsec:spikes_n_bursts}}

\begin{figure}[t]
\centering{}\includegraphics[width=1\columnwidth]{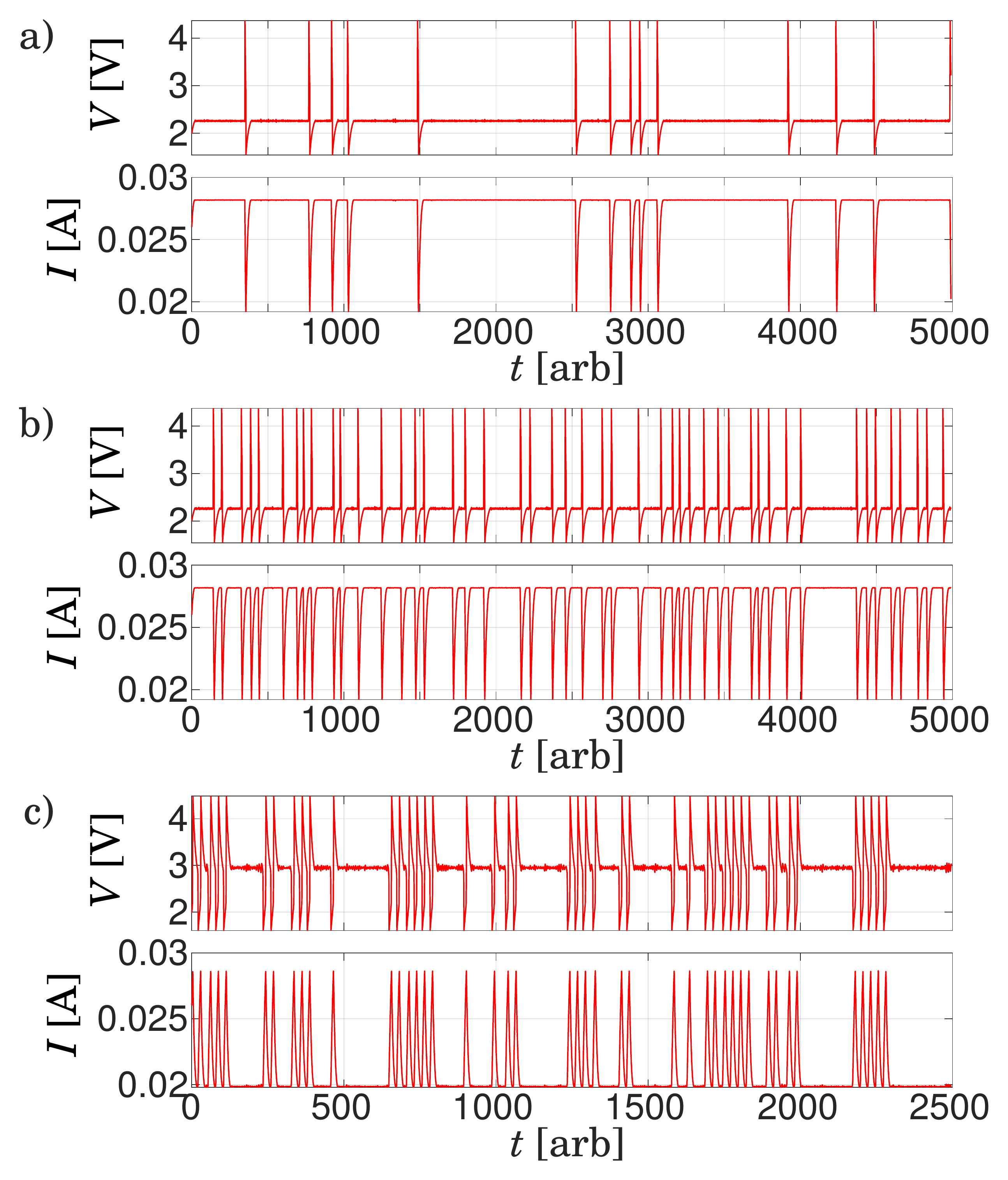}\caption{Numerical traces of output pulses randomly generated by perturbing
with additive white noise at $R=0.01\,\Omega$ and $\mu=0.001\,\Omega^{-1}$.
The system is biased in the first PDC region in panels \emph{a} and
\emph{b} and in the second PDC region in panel \emph{c}. The input
bias voltage and input noise intensity are: \emph{a}) $V_{0}=2.26\,\text{V}$,
$\eta=0.009\,\text{V}$, \emph{a}) $V_{0}=2.26\,\text{V}$, $\eta=0.011\,\text{V}$,
\emph{a}) $V_{0}=2.94\,\text{V}$, $\eta=0.014\,\text{V}$.\label{fig:spikes_n_bursts}}
\end{figure}

Adding noise into the system biased either in the first or second
PDC region is a way to perturb it in a permanent, random manner. For
instance, an electrical noise input can be incorporated in the model
as an additive white noise function $\xi(t)$ in equation \ref{eq:dVdt}.
Here, the stochastic process $\xi(t)$ satisfies $\left\langle \xi(t)\right\rangle =0$
and $\left\langle \xi(t)\xi(t')\right\rangle =\eta^{2}\delta(t'-t)$.
Some of these perturbations might be stronger than the excitability
threshold and will therefore trigger a spike. The higher the noise
intensity $\eta$, the more likely an above-threshold perturbation
is to occur and hence the more frequent the spikes are expected to
be, with the possibility to achieve coherence resonance behavior \citep{LGN-PR-04},
as no interval between consecutive spikes can be shorter than the
refractory time. Figure \ref{fig:spikes_n_bursts}\emph{a,b} shows
the system's response biased in the first PDC region, under the same
parameters $\left(R,\mu,V_{0}\right)$ and different levels of noise.
As expected, spikes are triggered more frequently as the noise intensity
is increased. Because of the strongly asymmetric character of the
I-V curve, there is a substantial qualitative difference between the
spikes generated with the circuit biased at the first and the second
PDC region, as in the latter case, the spikes tend to arise in bursts,
separated by the refractory time, as shown in figure \ref{fig:spikes_n_bursts}\emph{c}
and in agreement with the results of \citep{RJI-OE-13}.

\section{Bifurcation analysis  \label{sec:bifurcation}}

\begin{figure}[t]
\centering{}\includegraphics[width=1\columnwidth]{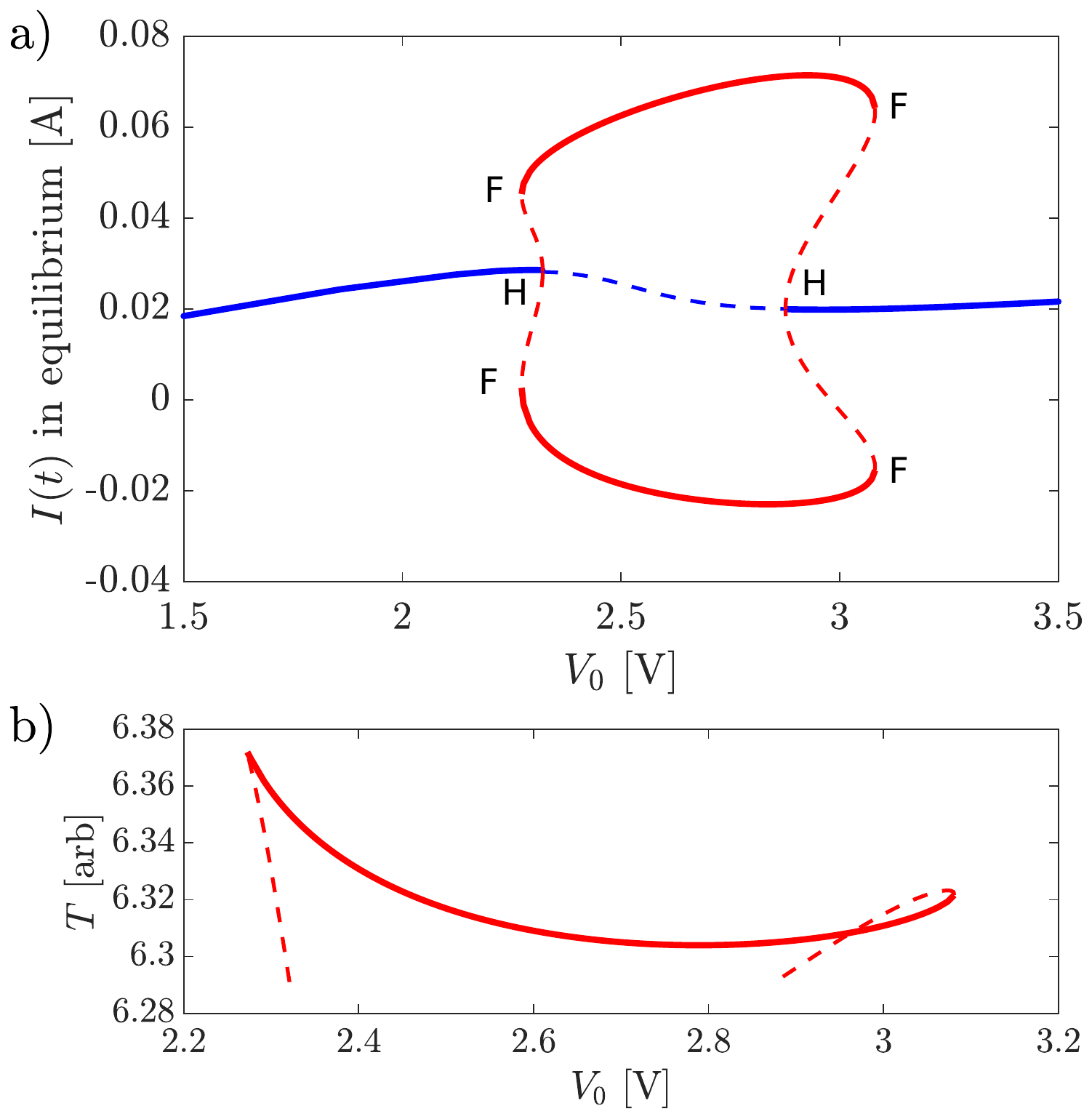}\caption{\emph{a}) Bifurcation diagram illustrating the $I$-coordinate of
the fixed point (in blue) and the $I$-maximum and minimum of the
limit cycle (in red) versus the input bias voltage $V_{0}$ for $R=1\,\Omega$
and $\mu=0.05\,\Omega^{-1}$. Solid lines represent stable solutions
and dashed lines represent unstable solutions. AH bifurcations (H)
and limit cycle folds (F) are also shown. \emph{b}) Evolution of the
period $T$ of the limit cycle along the branch.\label{fig:bifurcation_diagram_smooth}}
\end{figure}

\begin{figure}[t]
\centering{}\includegraphics[width=1\columnwidth]{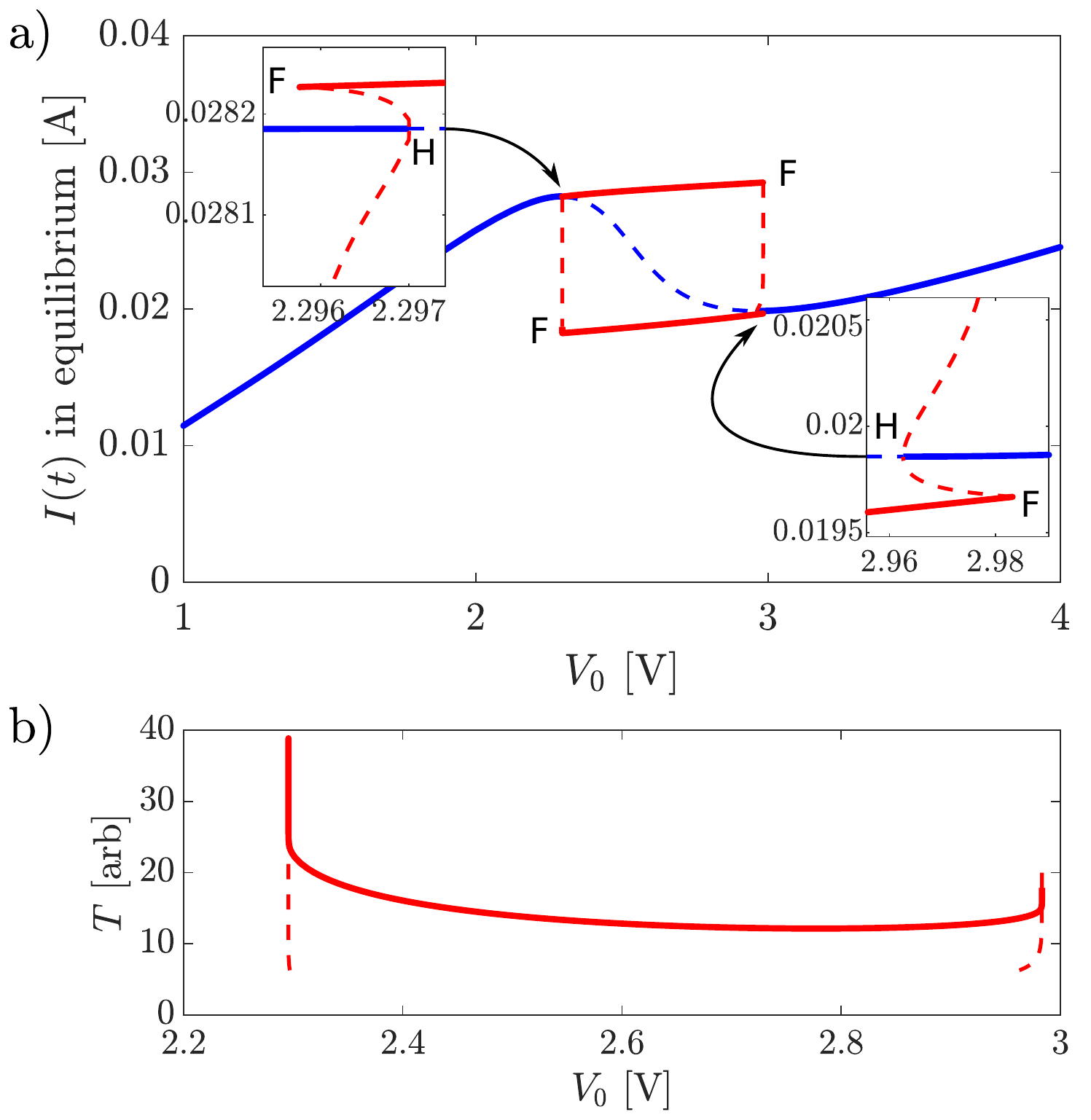}\caption{\emph{a}) Bifurcation diagram illustrating the $I$-coordinate of
the fixed point (in blue) and the $I$-maximum and minimum of the
limit cycle (in red) versus the input bias voltage $V_{0}$ for $R=1\,\Omega$
and $\mu=0.002\,\Omega^{-1}$. Solid lines represent stable solutions
and dashed lines represent unstable solutions. The insets show a zooming
in the proximity each AH bifurcation (H). The limit cycle folds (F)
are also shown. \emph{b}) Evolution of the period $T$ of the limit
cycle along the branch.\label{fig:bifurcation_diagram_stiff}}
\end{figure}
We analyze in this section the response of the system (i.e., stable
and unstable solutions in equilibrium) in terms of the free parameters
$R$, $\mu$ and $V_{0}$. In particular, we are interested in understanding
which parameter combination leads to an effective sigle-event excitable
spike generator versus a bursting generator. Two cases will be discussed
separately, depending on whether the intrinsic resistance $R$ is
under or over the critical value $R_{C}$. If $R<R_{C}$ (section
\ref{sec:R_small}), the nullclines intersect in a unique point and
therefore, the regions defined by the equilibrium responses and the
bifurcations that separate them are rather simple to describe. Indeed,
the geometry of these regions is not substantially affected by the
value of the resistance. The fixed point may change stability via
an AH bifurcation, which also gives rise to a limit cycle. The limit
cycle is stable or unstable depending on whether the bifurcation is
supercritical or subcritical. In the latter case, the unstable limit
cycle becomes stable via a fold bifurcation. On the other hand, if
$R>R_{C}$ (section \ref{sec:R_large}), there may be up to three
fixed points, which may arise or vanish via a saddle-node bifurcation.
This opens the possibility of limit cycles surrounding one fixed point
or all of them, as well as the possibility of a limit cycle colliding
a saddle point in a homoclinic bifurcation. All this makes the range
of responses and transitions more rich and complex to describe in
this case.

The branches of periodic and homoclinic solutions as well as the bifurcations
relating to those have been computed by using DDE-BifTool \citep{DDEBT}
package (version 3.1.1). The fixed point branches and AH bifurcations
have been computed analytically through curve parametrization. The
saddle-node bifurcations have been evaluated numerically.

\subsection{Case $R<R_{C}$\label{sec:R_small}}

\subsubsection{Fixed point and limit cycle branches\label{sec:bifurcation_diagrams}}

\begin{figure*}[t]
\centering{}\includegraphics[width=1\textwidth]{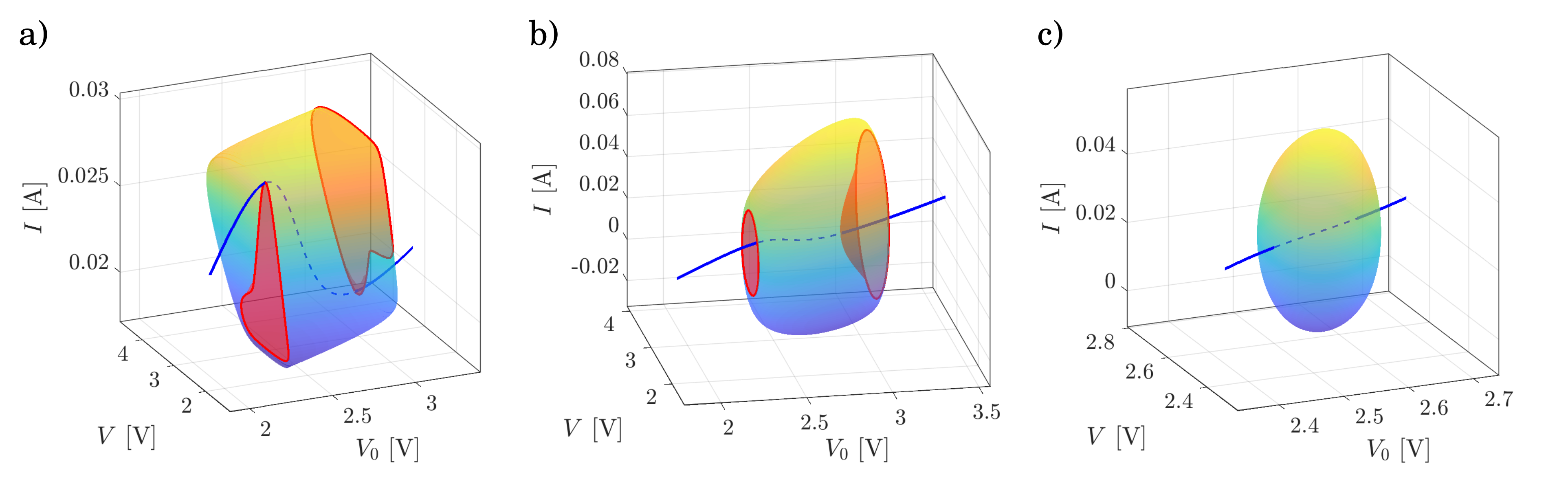}\caption{Three-dimensional bifurcation diagram illustrating the stable and
unstable steady stable branches, solid and dashed blue lines, respectively,
the stable limit cycle branch (colored gradient surface), the unstable
limit cycle branch (red surface) and limit cycle fold (red curves)
versus the input bias voltage $V_{0}$ for $R=1\,\Omega$ and different
values of $\mu$: a) $\mu=0.002\,\Omega^{-1}$, b) $\mu=0.05\,\Omega^{-1}$,
c) $\mu=0.14\,\Omega^{-1}$.\label{fig:bifurcation_3D_low_R}}
\end{figure*}
The goal in this section is to trace the evolution of the equilibrium
solutions in terms of the input bias voltage $V_{0}$ while the parameters
$R$ and $\mu$ are fixed. As discussed in part \ref{sec:theoretical_model},
calculation of the fixed point requires in principle solving the system
of equations (\ref{eq:I-V_nullcline},\ref{eq:load_line}), which
cannot be done analytically given the complexity of the function $f(V)$.
However, a branch of fixed points can be characterized as a parametric
curve; instead of tracking the evolution of the fixed point (namely,
its coordinates $V$ and $I$) in terms of $V_{0}$, Eqs.~(\ref{eq:I-V_nullcline},\ref{eq:load_line})
may be re-written as,

\begin{eqnarray}
V_{0} & = & V+RI,\label{eq:FP_param_V0}\\
I & = & f(V),\label{eq:FP_param_I}
\end{eqnarray}
thus expressing $V_{0}$ and $I$ in terms of $V$, which serves as
the free parameter. This allows to continue the branch without the
need to solve any equations; neither analytically nor numerically.
This method was used to compute all the fixed point branches presented
in this work. The parametric curve $\left(V_{0}(V),I(V)\right)$ is
plotted in figures \ref{fig:bifurcation_diagram_smooth}(a) and \ref{fig:bifurcation_diagram_stiff}(a)
for $R=1\,\Omega$ and two different values of $\mu$.

Note how the fixed point parametrization does not depend on $\mu$.
The stability, however, does; The fixed points are unstable when either
$f'(V)<-\mu^{2}R$ or $f'(V)<-1/R$. The latter cannot happen since
$R<R_{C}$ (i.e., $f'(V)>-1/R$ always) but the former can. In fact,
under the assumption that $f(V)$ is monotonous in the NDC region,
its derivative is then a basin-shaped function with a unique minimal
value $f'_{\text{MIN}}=-1/R_{C}<0$. Thus, provided that $\mu{{}^2}<1/(RR_{C})$,
the equation $f'(V)=-\mu^{2}R$ has exactly two solutions. Let us
call these solutions $V^{\left(1\right)}$ and $V^{\left(2\right)}$.
The fixed point is unstable if the input bias voltage is between the
values $V_{0}^{\left(1,2\right)}=V^{\left(1,2\right)}+Rf(V^{\left(1,2\right)})$
and it is stable elsewhere. These transitions are candidates of AH
bifurcations. Indeed, substitution of $f'(V)=-\mu^{2}R$ in equation
\ref{eq:eigenvalues} leads to, $\lambda_{\pm}=\pm i\sqrt{1-\left(\mu R\right)^{2}}$,
which is a purely imaginary number (provided that $R<R_{C}$ and $\mu{{}^2}<1/(RR_{C})$,
then $\left(\mu R\right)^{2}<1$). On the other hand, if $\mu{{}^2}\geq1/(RR_{C})$,
the fixed point is stable regardless of $V_{0}$.

The limit cycle branches were computed with DDE-BifTool. For sufficiently
large values of $\mu$, the periodic branch arises subcritically from
the fixed point at $V_{0}^{L}=2.231\,\text{V}$ and $V_{0}^{R}=2.876\,\text{V}$,
see Figure \ref{fig:bifurcation_diagram_smooth}(a) obtained with
$R=1\,\Omega$ and $\mu=0.05\,\Omega^{-1}$. For this relatively large
value of $\mu$, the dynamics is smooth and so are the solution branches.
There are two ranges of input bias where the stable fixed point and
the stable limit cycle coexist. These ranges have a width of about
0.05 V at the left side and 0.2 V at the right side. Figure \ref{fig:bifurcation_diagram_smooth}(b)
shows the limit cycle period in terms of $V_{0}$. The period is locally
maximized at the proximity of both fold bifurcations and it has a
plateau in the stable part of the branch, at about $V_{0}=2.8\,\text{V}$.
The solutions along this branch have a smooth, (quasi) sinusoidal
profile.

Figure \ref{fig:bifurcation_diagram_stiff}(a) illustrates the bifurcation
diagram for the same value of $R=1\,\Omega$, but a much smaller stiffness
coefficient $\mu=0.002\,\Omega^{-1}$ allowing for slow-fast dynamics.
While the AH bifurcations points are essentially not displaced ($V_{0}^{L}=2.297\,\text{V}$
and $V_{0}^{R}=2.961\,\text{V}$), the branch of periodic solutions
changes dramatically; in the proximity of the AH bifurcation points,
the amplitude of the limit cycle increases explosively. A zooming
in both regions show that the bifurcations are still sub-critical,
whilst the ranges of bistability are much more narrow (in the order
of 0.001~V at the left side and 0.01~V at the right side). On both
sides, the unstable limit cycle folds and becomes stable. Solutions
along the stable part of the branch are very similar to those in figure
\ref{fig:slowfast}(b), with stages of low and fast dynamics that
can be clearly distinguished, and their amplitude and shape do not
change substantially with $V_{0}$. Figure \ref{fig:bifurcation_diagram_stiff}(b)
shows the period along the branch. The results are qualitatively similar
to those in figure \ref{fig:bifurcation_diagram_smooth}(b); the period
is maximal at the proximity of the AH bifurcations with a plateau
in between, where the limit cycle is stable, although the increasing
of the period towards the maximal values is very sharp. There is an
important quantitative difference, however, as the period is one order
of magnitude larger than that computed for $\mu=0.05\,\Omega^{-1}$.

The bifurcation diagrams from figures \ref{fig:bifurcation_diagram_stiff}(a)
and \ref{fig:bifurcation_diagram_smooth}(a) are shown again in perspective
in figure \ref{fig:bifurcation_3D_low_R} (panels \emph{a} and \emph{b},
respectively), where both coordinates $V$ and $I$ of the equilibrium
solutions are plotted versus the parameter $V_{0}$. A new bifurcation
diagram is included where $\mu=0.14\,\Omega^{-1}$ (panel \emph{c}).
We already know that for $\mu=0.002\,\Omega^{-1}$, the dynamics is
stiff, the limit cycle folds are sharp and the bistability ranges
are very narrow. As $\mu$ is increased, the limit cycle branch becomes
smoother and the bistability ranges widen, particularly the right
one, which is always wider than the left one. For values of $\mu$
above 0.05~$\Omega^{-1}$ however, this tendency has reversed, and
the bistability ranges become narrow again, until they eventually
vanish. At $\mu=0.14\,\Omega^{-1}$, the AH bifurcations have become
supercritical and there is no unstable periodic solutions nor periodic
solution folds (panel \emph{c}). In regards to the AH points, they
become closer as $\mu$ increases and they eventually coalesce at
$\mu<1/\sqrt{RR_{C}}\approx0.1703\,\Omega^{-1}$. Consequently, the
periodic solutions also vanish.

\subsubsection{Evolution of the bifurcations in the space of parameters\label{subsec:muanalysis}}

\begin{figure}[t]
\centering{}\includegraphics[width=1\columnwidth]{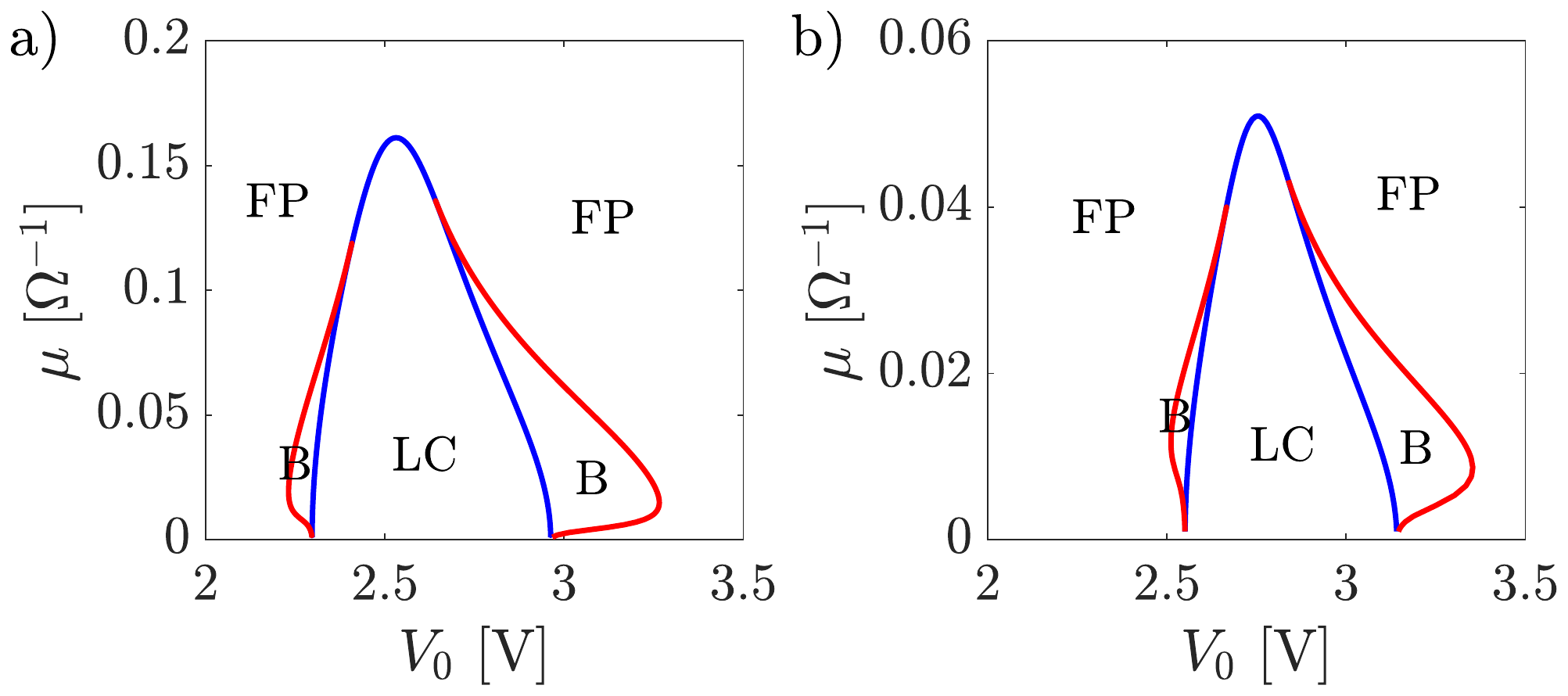}\caption{Andronov-Hopf bifurcation branches (blue line) and limit cycle fold
branches (red line) on the $\left(V_{0},\mu\right)$ plane for $R=1\,\Omega$
(\emph{a}) and $R=10\,\Omega$ (\emph{b}). There is a unique stable
fixed point (FP) outside the region delimited by the AH branch which
is unstable inside. In this region, there is a stable limit cycle
(LC). In the regions between the AH branch and the limit cycle fold
branch, the system is bistable (B) in the sense that the stable fixed
point and the stable limit cycle coexist.\label{fig:branches_low_R}}
\end{figure}

In this section, the AH bifurcation and limit cycle fold branches
are characterized as curves in the $\left(V_{0},\mu\right)$ plane
while $R$ is fixed at different values under $R_{C}$. The AH points
satisfy the equation $f'(V)=-\mu^{2}R$. This, along with equations
\ref{eq:I-V_nullcline} and \ref{eq:load_line} are used to parametrize
the AH branch in terms of $V$,

\begin{eqnarray}
V_{0} & = & V+Rf(V),\label{eq:hopf_param_V0}\\
\mu & = & \sqrt{-\frac{f'(V)}{R}},\label{eq:hopf_param_mu}
\end{eqnarray}

where $V$ is in the NDC region (so $\mu(V)\in\mathbb{R}$). The latter
reaffirms that AH bifurcations can only occur in the NDC region. Figure
\ref{fig:branches_low_R} show the AH branch for different resistances
under $R_{C}$. In general, the branch defines a uniquely evaluated,
hill-shaped function of $\mu$ versus $V_{0}$. The maximal value
of $\mu$ at the top of the branch is $\mu_{\text{MAX}}=1/\text{\ensuremath{\sqrt{RR_{C}}}}$,
where the left and right AH points collide. Above this point, there
is no unstable fixed points nor limit cycles at all.

The limit cycle fold branches were numerically computed with DDE-BifTool
and they are also included in figure \ref{fig:branches_low_R}. As
$\mu$ approaches zero, the fold branches approach the ends of the
AH branch asymptotically. As $\mu$ increases, the fold points move
away from the AH branch and the bistability ranges (coexistence of
an attractor with a stable limit cycle) broaden, until they reach
a plateau and the fold branches approach the AH branch tangentially
until they collide with it close to the top. Consequently, the AH
bifurcation transitions from sub-critical to supercritical. In general,
the bistability range at the right of the AH branch is broader than
the one at the left. This will be discussed in more detail in section
\ref{subsec:suitable_parameters}.

The above analysis is consistent with figures \ref{fig:bifurcation_diagram_smooth},
\ref{fig:bifurcation_diagram_stiff} and \ref{fig:bifurcation_3D_low_R},
and it is valid for all positive resistances under $R_{C}$ as the
branches retain their geometric characteristics. On the quantitative
basis, however, the branches shift rightward and become smaller with
increasing \emph{$R$}, as $\mu_{\text{MAX}}$ decreases with the
inverse of the square root of the resistance.

\subsection{Case $R>R_{C}$\label{sec:R_large}}

\subsubsection{Fixed point and limit cycle branches\label{sec:solution_branches_high_R}}

\begin{figure*}[t]
\centering{}\includegraphics[width=1\textwidth]{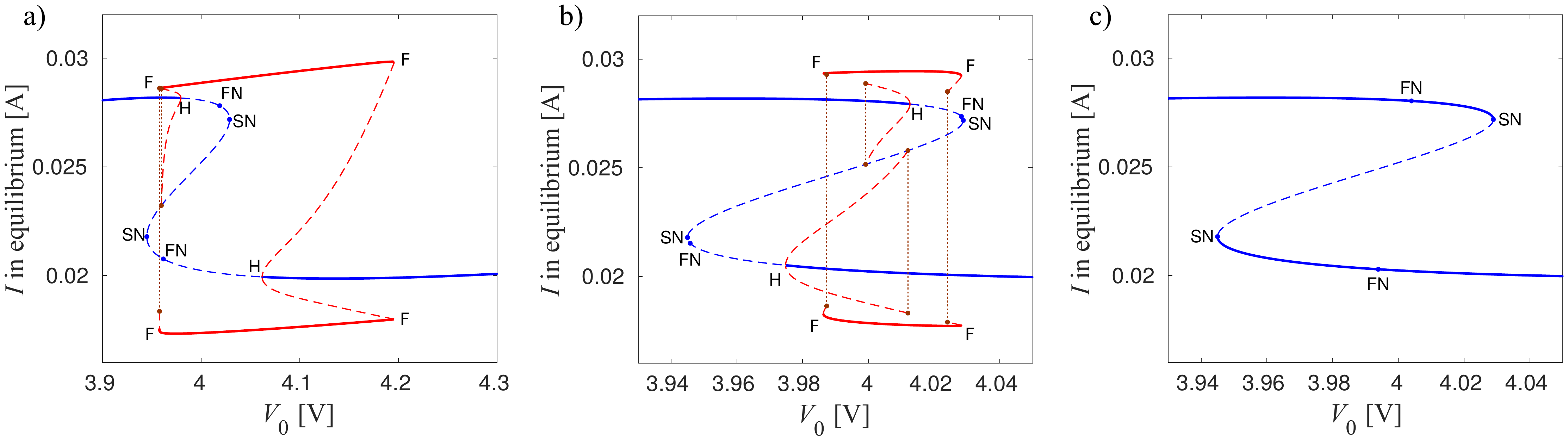}\caption{Bifurcation diagram illustrating the $I$-coordinate of the fixed
point (blue lines) as well as the $I$-maximum and minimum of the
limit cycle (red lines) and homoclinic curves (brown dots) versus
the input bias voltage $V_{0}$ for $R=60\,\Omega$ and different
values of $\mu$: \emph{a}) $\mu=0.006\,\Omega^{-1}$, \emph{b}) $\mu=0.0115\,\Omega^{-1}$,
\emph{c}) $\mu=0.03\,\Omega^{-1}$. Solid lines represent stable solutions
and dashed lines represent unstable solutions. Andronov-Hopf bifurcations
(H), limit cycle folds (F), saddle-node bifurcations (SN) and focus-node
transitions (FN) are also illustrated. As $\mu$ increases, the AH
bifurcations and focus-node transitions approach the saddle-node bifurcation,
and a second pair of homoclinic curves arise. For $\mu>1/R$, there
is no AH bifurcation and no periodic nor homoclinic solutions.\label{fig:bifurcation_hcli}}
\end{figure*}

If $R>R_{C}$ the system may exhibit up to three fixed points since
the nullclines may now have more than one intersection, which depends
specifically on the position coefficient of the load line, i.e., the
input bias voltage $V_{0}$. Consequently, the fixed point branch
is not uniquely evaluated in terms of $V_{0}$ but it folds backwards
and then forwards again, as illustrated in figure \ref{fig:bifurcation_hcli}.
In this regime the RTD may behave as a memory \citep{UTM-IEICE-99}.
The folds occur when the load line intersects the current-voltage
characteristic tangentially and induce saddle-node bifurcations. This
can be corroborated by computing the eigenvalues at the folds; Tangential
intersection implies $f'(V)=-1/R$. Substitution of the latter in
equation \ref{eq:eigenvalues} reduces the eigenvalues to $\lambda_{+}=0$
and $\lambda_{-}=\tfrac{1}{\mu R}-\mu R$ at both folds. If $\mu<1/R$,
the first eigenvalue is positive, suggesting that as the fixed point
crosses the saddle-node bifurcation, one eigenvalue remains positive
while the other changes sign, thus transitioning from an unstable
node to a saddle point. All the latter implies that the fixed point
transitions from an unstable focus to an unstable node somewhere in
between the AH bifurcation and the saddle-node. The process is reversed
as we move along the branch; the saddle point becomes an unstable
node again, then an unstable focus and finally, a stable focus at
the second AH bifurcation. All these transitions are shown in figure
\ref{fig:bifurcation_hcli}, panels \emph{a} and \emph{b}. As $\mu$
approaches $1/R$, the transitions become closer and coalesce at each
fold. On the other hand, if $\mu>1/R$, the non-zero eigenvalue is
negative; as the fixed point passes along the fixed point branch (and
thus, the saddle-nodes), it transitions from a stable node to a saddle
point and then back into a stable node (figure \ref{fig:bifurcation_hcli},
panel \emph{c}). Thus, there is no AH bifurcation when $\mu>1/R$.
From all the above it is concluded that, whenever the system exhibits
three fixed equilibrium points, the middle one is always a saddle.
Finally, by using the parametric form of the fixed point branch (eqs
(\ref{eq:FP_param_V0},\ref{eq:FP_param_I})), we find that the bias
voltages for which the saddle-nodes occur are $V_{0}=V_{\text{SN1}}+Rf(V_{\text{SN1}})$
and $V_{0}=V_{\text{SN2}}+Rf(V_{\text{SN2}})$, where $V_{\text{SN1}}$and
$V_{\text{SN2}}$ are the solutions of $f'(V)=-1/R$.

\begin{figure}[t]
\centering{}\includegraphics[width=1\columnwidth]{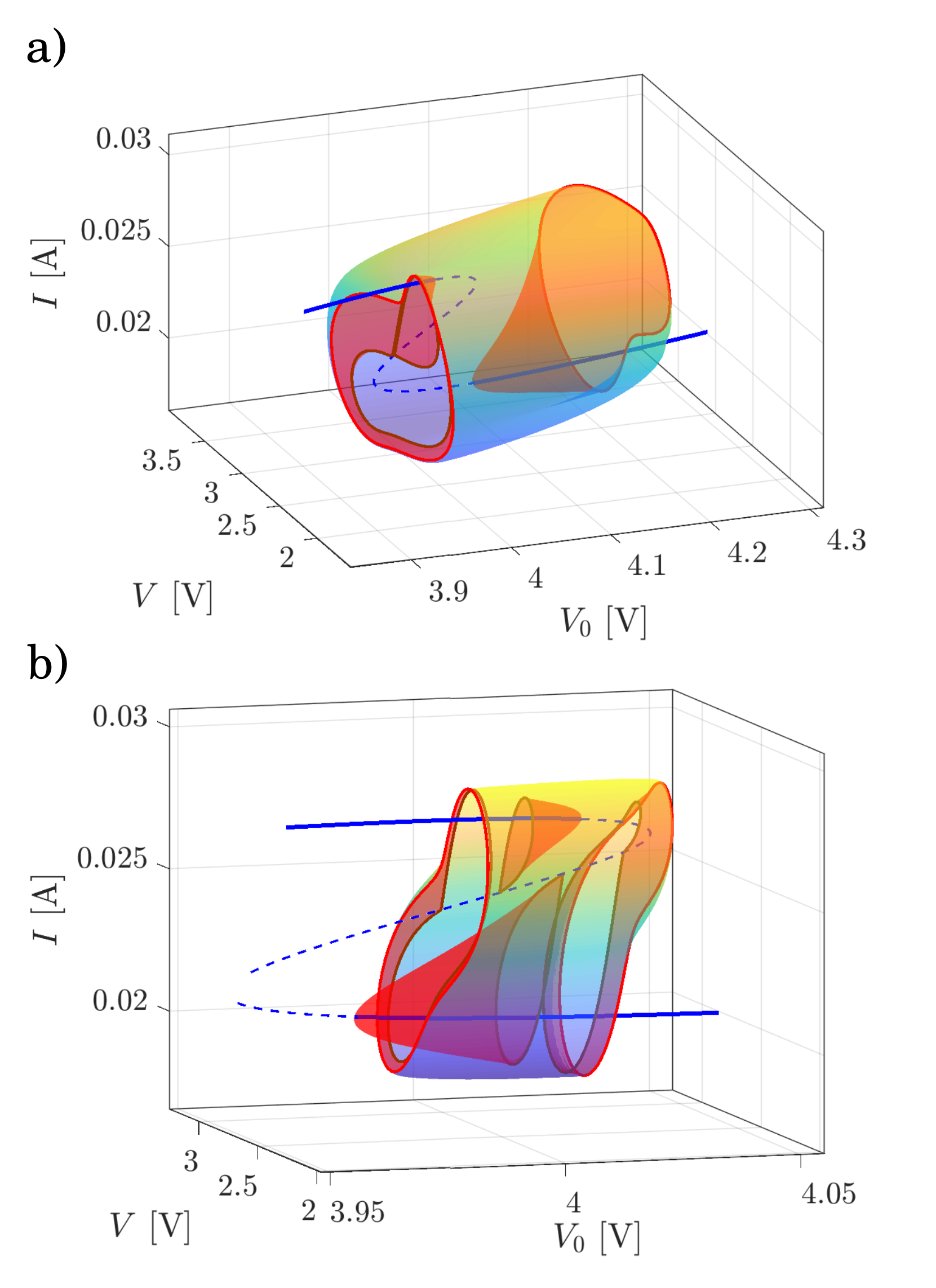}\caption{Three-dimensional bifurcation diagram illustrating the stable point
branch (solid blue line), unstable point branch (dashed blue line),
stable limit cycle branch (color gradient surface), unstable limit
cycle branch (red surface), limit cycle fold (red curves) and homoclinic
solutions (brown curves) versus the input bias voltage $V_{0}$ for
$R=60\,\Omega$ and different values of $\mu$: a) $\mu=0.006\,\Omega^{-1}$,
b) $\mu=0.0115\,\Omega^{-1}$.\label{fig:bifurcation_3D_high_R}}
\end{figure}

The existence of a saddle fixed point in the system allows for the
possibility of a limit cycle colliding with it and thus becoming a
homoclinic curve, which separates the limit cycle branch into two
or more branches. A tiny separation can be appreciated in figure \ref{fig:bifurcation_hcli}(a).
From the upper AH bifurcation, $V_{0}\approx3.98\,\text{V}$, an unstable
limit cycle arises, which surrounds the upper attractor. The branch
continues until the lower part of the limit cycle coalesces with the
saddle point and becomes a homoclinic curve, at about $V_{0}\approx3.96\,\text{V}$.
At a slightly smaller input bias, a new unstable limit cycle arises
from a homoclinic curve that surrounds all the fixed points. This
limit cycle branch then folds, becomes stable and extends until $V_{0}\approx4.196\,\text{V}$,
where it folds again and becomes unstable. The unstable limit cycle
collides with the fixed point in the lower AH bifurcation at $V_{0}\approx4.062\,\text{V}$.
As $\mu$ increases (while $R$ is kept fixed), the right fold of
the limit cycle branch approaches -and enters- the range defined by
the saddle-nodes. Consequently, the right unstable limit cycle branch
collides with the saddle fixed point branch, which produces two new
homoclinic curves, as shown in figure \ref{fig:bifurcation_hcli}.\emph{b}.
As $\mu$ is further increased, the limit cycles and homoclinic curves
vanish (figure \ref{fig:bifurcation_hcli}.\emph{c}) by mechanisms
that will be explained in section \ref{subsec:evolution_high_R}.

\begin{figure}[t]
\begin{centering}
\includegraphics[width=1\columnwidth]{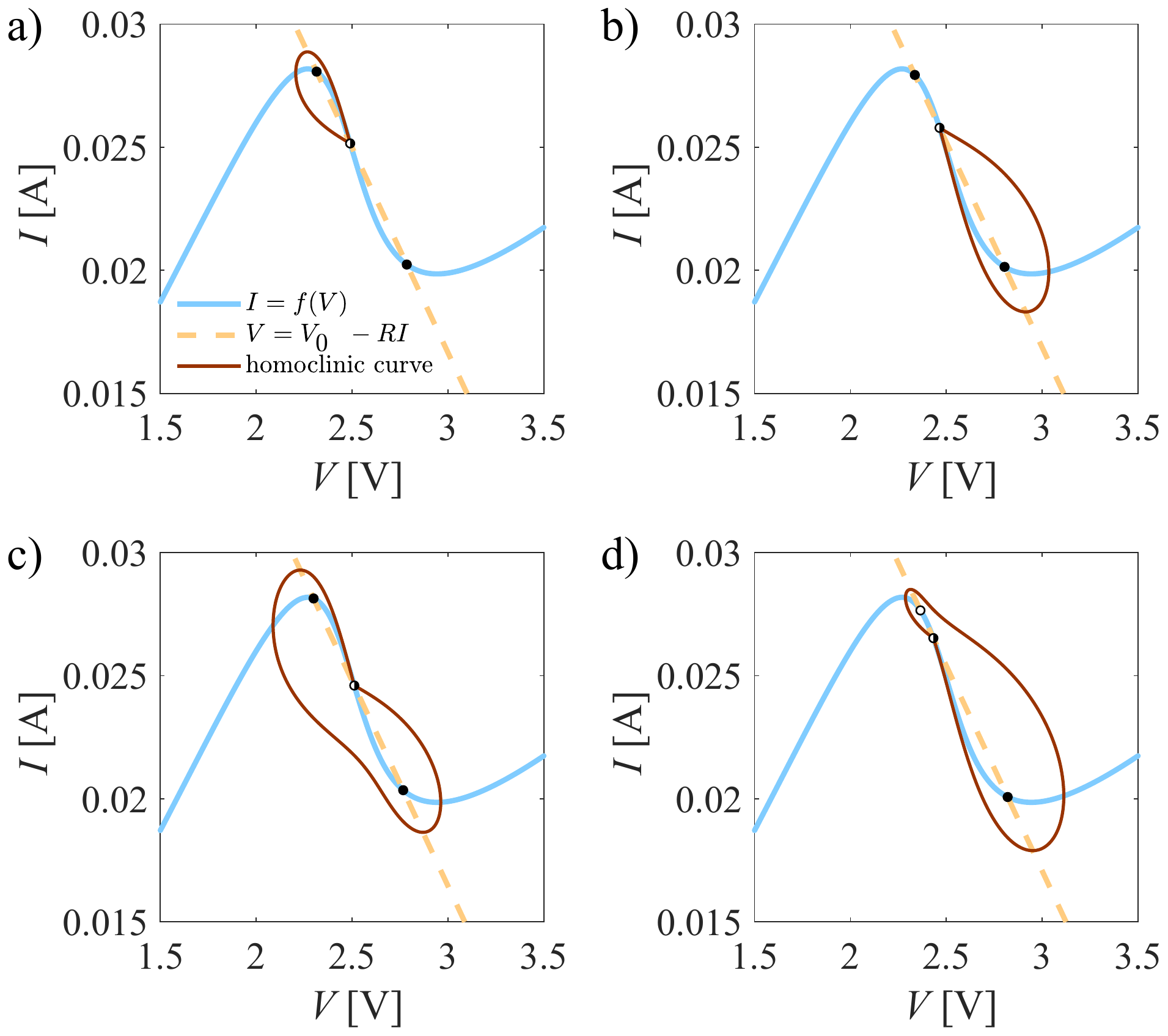}\caption{Homoclinic curves shown in figures \ref{fig:bifurcation_hcli}.\emph{b}
and \ref{fig:bifurcation_3D_high_R}.\emph{b,} arising for parameters
$R=60\,\Omega$, $\mu=0.0115\,\Omega^{-1}$ and different values of
$V_{0}$: \emph{a}) Type-1 homoclinic, $V_{0}=3.999\,\text{V}$\emph{.
b}) Type-2 homoclinic, $V_{0}=4.012\,\text{V}$\emph{. c}) Type-3
homoclinic, $V_{0}=3.987\,\text{V}$\emph{. d}) Type-4 Homoclinic,
$V_{0}=4.024\,\text{V}$.\label{fig:homoclinics}}
\par\end{centering}
\end{figure}

Figure \ref{fig:bifurcation_3D_high_R} shows the bifurcation diagrams
from \ref{fig:bifurcation_hcli}, panels \emph{a} and \emph{b}, in
perspective, including both coordinates $V$ and $I$. The two homoclinic
curves in the middle surround the upper attractor and the lower attractor,
respectively. These homoclinics are illustrated in figure \ref{fig:homoclinics},
panels \emph{a} and \emph{b}. We refer to these curves as type-1 homoclinic
and type-2 homoclinic. The remaining two outer homoclinics surround
all three fixed points as illustrated in figure \ref{fig:homoclinics},
panels \emph{c} and \emph{d}. We refer to these curves as type-3 homoclinic
and type-4 homoclinic, respectively.

\subsubsection{Evolution of the bifurcations in the space of parameters\label{subsec:evolution_high_R}}

\begin{figure}[t]
\begin{centering}
\includegraphics[width=1\columnwidth]{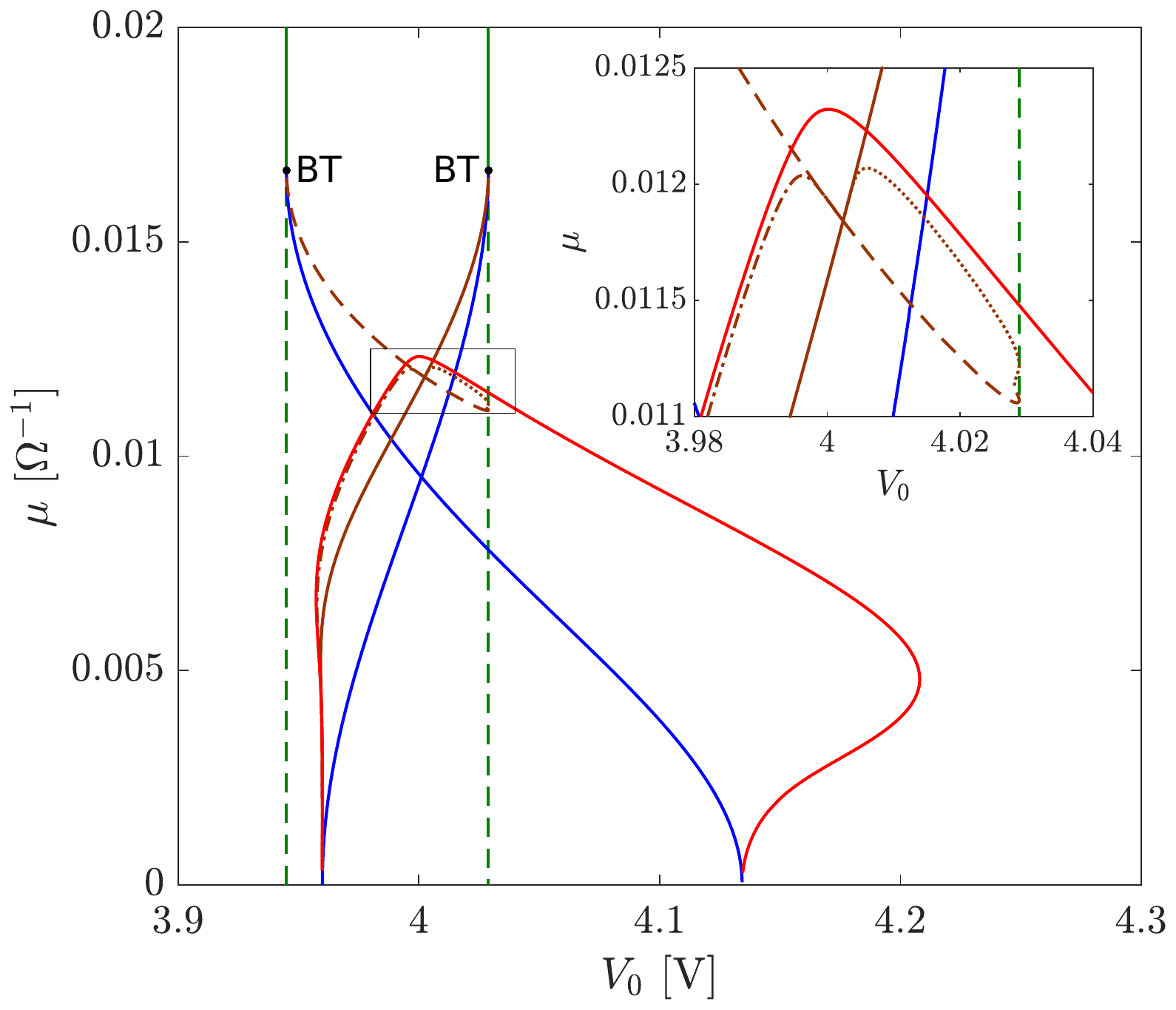}\caption{ Bifurcation branches represented as curves on the on
the $\left(V_{0},\mu\right)$ plane for $R=60\,\Omega$: Andronov-Hopf
bifurcations (blue line), limit cycle fold bifurcation (red line),
saddle-stable node bifurcations (solid green line), saddle-unstable
node bifurcations (dashed green line), type-1 homoclinic bifurcation
(solid brown line), type-2 homoclinic bifurcation (dashed brown line),
type-3 homoclinic bifurcation (dashed-dotted brown line), type-4 homoclinic
bifurcation (dotted brown line), Bogdanov-Takens bifurcations (BT).
The inset zooms into the black square in the figure, showing the homoclinic
bifurcations in detail. \label{fig:branches_high_R}}
\par\end{centering}
\end{figure}

Figure \ref{fig:branches_high_R} illustrates the bifurcation branches
as curves in the $\left(V_{0},\mu\right)$ plane, for a fixed parameter
$R=60\,\Omega$. The two saddle-node bifurcations define two parallel,
vertical branches since, as already discussed in section \ref{sec:solution_branches_high_R},
their positions in the bifurcation diagram do not depend on $\mu$
and are given by $V_{0}=V_{\text{SN1}}+Rf(V_{\text{SN1}})$ and $V_{0}=V_{\text{SN2}}+Rf(V_{\text{SN2}})$,
where $f'(V_{\text{SN1}})=f'(V_{\text{SN2}})=-1/R$. For $R=60\,\Omega$,
the positions of the saddle-nodes are estimated at $V_{0}\approx4.029\,\text{V}$
and $V_{0}\approx3.945\,\text{V}$. It was also discussed in section
\ref{sec:solution_branches_high_R} that, above $\mu=1/R$, the saddle-nodes
correspond to transitions from a stable node to a saddle, while below
$\mu=1/R$, they correspond to transitions from an unstable node to
a saddle. This is represented in figure \ref{fig:branches_high_R}
as solid and dashed green lines, respectively.

Figure \ref{fig:branches_high_R} also shows that when $R>R_{C}$,
there is not one AH bifurcation branch but two of them. This happens
because, since the fixed point branch is not a uniquely evaluated
curve, the upper and lower AH bifurcations do not coalesce with one
another but switch sides instead, at $\mu\approx0.0095\,\Omega^{-1}$
(this can be appreciated in figure \ref{fig:bifurcation_hcli}.\emph{b},
where the upper AH point is now at the right side and the lower AH
point, at the left side). As $\mu$ is further increased, the lower
and upper AH branches approach the left and right saddle-nodes, respectively,
and they eventually coalesce at $\mu=1/R\approx0.0167$. 

From the analytical point of view, the parametric form of the AH bifurcation
branch given by Eqs~(\ref{eq:hopf_param_V0},\ref{eq:hopf_param_mu})
is still valid. However, an additional restriction applies when $R>R_{C}$.
The eigenvalues at the AH points were computed as $\lambda_{\pm}=\pm i\sqrt{1-\left(\mu R\right)^{2}}$,
which are meant to be purely imaginary; thus, $\mu<1/R$. Substitution
of parametric equation \ref{eq:hopf_param_mu} in the latter inequality
leads to $f'(V)>1/R$. This restriction, together with the restriction
$f'(V)<0$ required by equation \ref{eq:hopf_param_mu}, imply that
the curve parameter $V$ must be in the NDC region, but not in the
interval $\left[V_{\text{SN1}},V_{\text{SN2}}\right]$. In conclusion,
the AH branch being split into two results from a discontinuity in
the domain of the parametric form.

The limit cycle fold and homoclinic branches were continued with DDE-BifTool.
The branches are illustrated in figure \ref{fig:branches_high_R}.
When $R>R_{C}$, there is a single limit cycle fold branch instead
of two. This happens because the homoclinic bifurcations disconnect
the stable part of the limit cycle branch and thus the folds do not
coalesce with the AH points. Instead, the folds become closer as $\mu$
increases until they coalesce at the top of the fold branch at $\mu\approx0.0123\,\Omega^{-1}$.
Above this point, there is no stable limit cycles anymore. The type-1
and type-3 homoclinic branches arise at $\mu\approx0.0111\,\Omega^{-1}$,
as the left unstable limit cycle branch coalesces with the saddle
point branch in the proximity to the left saddle-node. On the other
hand, the type-2 and type-4 homoclinic branches arise at the bottom
of the $\left(V_{0},\mu\right)$ plane, together with the upper AH
and left limit cycle fold branches. The type-3 and type-4 homoclinic
branches remain close to each side of the limit cycle fold branch.
At $\mu\approx0.012\,\Omega^{-1}$, they fold down and at $\mu\approx0.0118\,\Omega^{-1}$,
they coalesce with the type-1 and type-2 homoclinics and with each
other. The type-3 and type-4 homoclinics vanish, while the type-1
and type-2 homoclinics simply switch sides, as they involve opposite
eigendirections of the saddle point. At $\mu=1/R\approx0.0167$, the
type-1 homoclinic branch coalesce with the upper AH branch and with
the right saddle-node branch. This triple coalescence is known as
a Bogdanov-Takens (BT) bifurcation and has codimension 2 \citep{Izhikevich}.
Also at $\mu=1/R$, the type-2 homoclinic branch coalesce with the
lower AH branch and the left saddle-node branch in another BT bifurcation.

\begin{figure}[!t]
\centering{}\includegraphics[width=0.8\columnwidth]{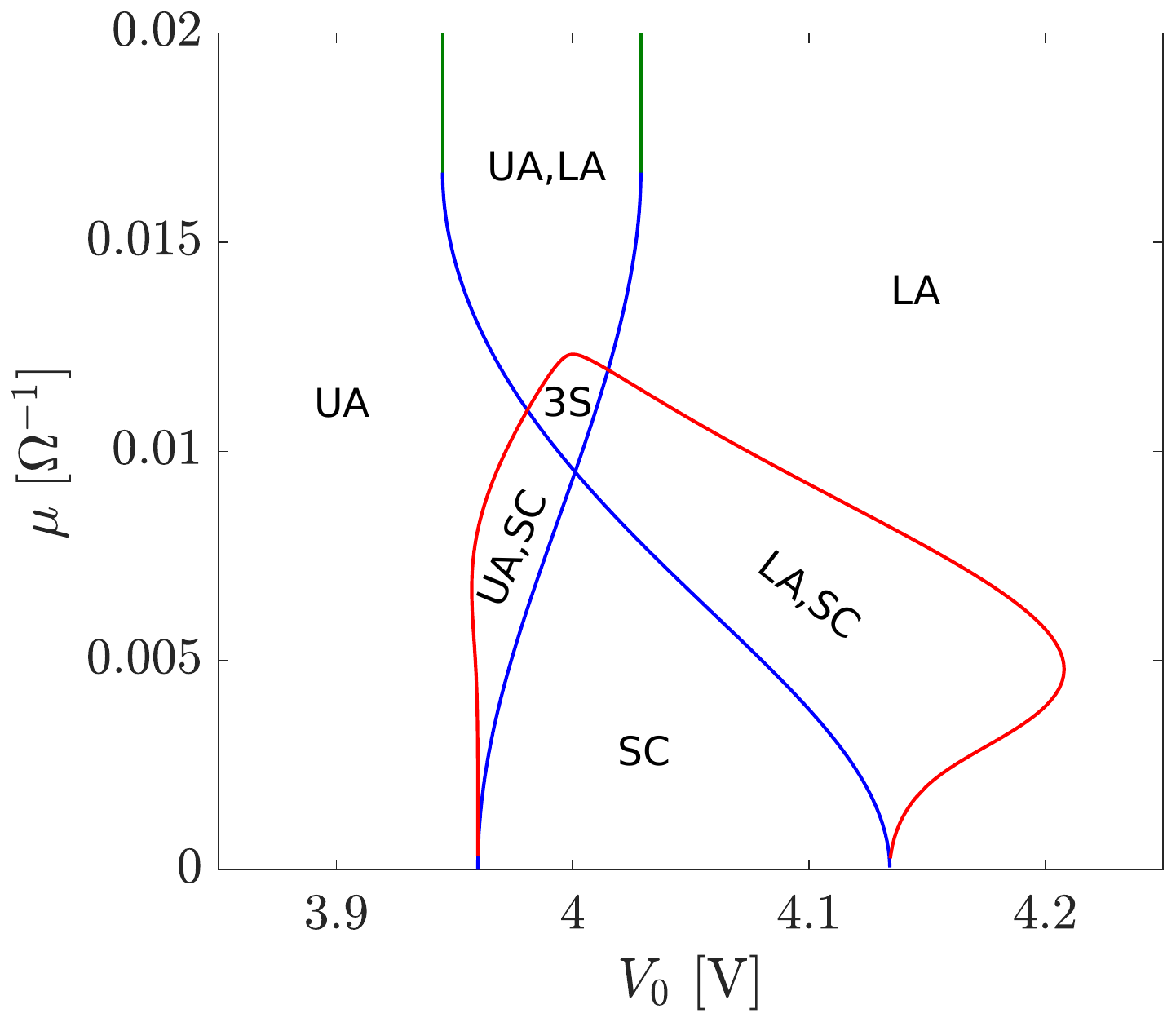}\caption{(color online) Stable solutions and the bifurcations that delimit
them on the $\left(V_{0},\mu\right)$ plane for $R=60\,\Omega$. UA:
upper attractor point. LA: lower attractor point. SC: stable limit
cycle. 3S: coexistence of the three stable solutions.\label{fig:branches_high_R_SS}}
\end{figure}

The description of the stable solutions becomes more complex when
$R>R_{C}$, as there may be now two stable fixed points in addition
to the stable limit cycle, which in turns allows more multistabilities.
These mutistabilities are summarized Fig~\ref{fig:branches_high_R_SS},
which only includes the bifurcation branches beyond which the stable
solutions vanish or become unstable, i.e, the AH bifurcations, the
limit cycle fold and the saddle-node bifurcations above $\mu>1/R$.
The saddle-node and AH branches limit the upper attractor the the
left side of the figure, the lower saddle-node and AH branches limit
the lower attractor to the right side, and the limit cycle fold limit
the stable limit cycle inside of it. These regions intersect and produce
bistabilities as well as a small region at the center of the figure
where the three stable solutions coexist.

\subsection{Suitable parameters for spike generation\label{subsec:suitable_parameters}}

\begin{figure}[th]
\centering{}\includegraphics[width=0.9\columnwidth]{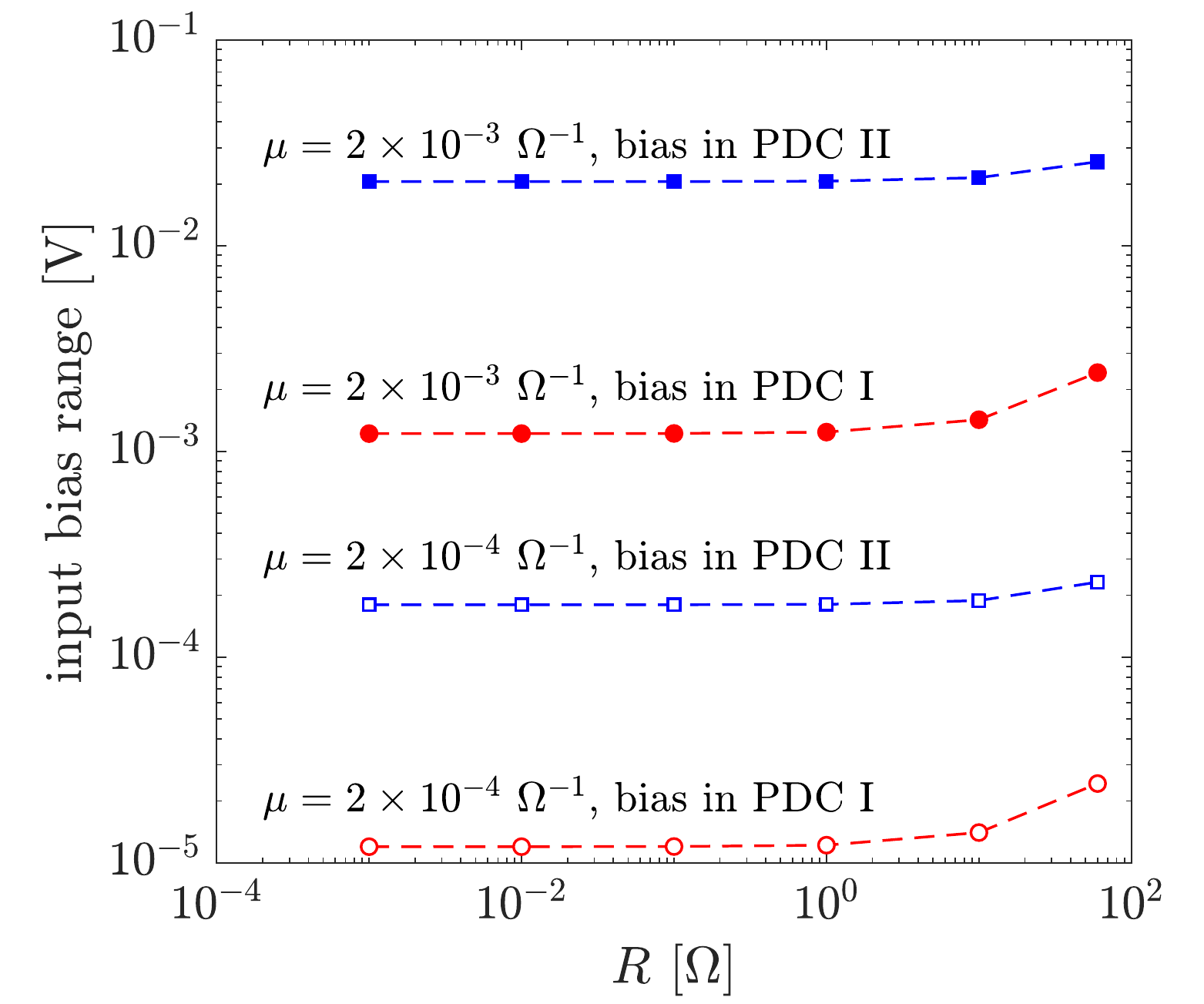}\caption{Input bias voltage range of bistability region in terms of the resistance
$R$ at the first and second PDC regions (red circles and blue squares,
respectively). The values of the stiffness coefficient are $\mu=0.0002\,\Omega^{-1}$
(empty symbols) and $\mu=0.002\,\Omega^{-1}$ (solid symbols).\label{fig:bias_range}}
\end{figure}

In regards to the interest in utilizing the RTD circuit theoretically
represented by equations (\ref{eq:dVdt},\ref{eq:dIdt}) as an excitable
spike generator, it has already been discussed that the stiffness
parameter $\mu$ needs to be small (no bigger than $\sim0.002\,\Omega^{-1}$)
for the stable limit cycle to produce spikes. As discussed in section
\ref{subsec:excitable}, the input bias voltage $V_{0}$ needs to
be tuned in such a way either the upper or the lower fixed point is
stable, but in the proximity of the region where there is a stable
limit cycle. For the parameters of the nonlinear function $f\left(V\right)$
chosen in this work and resistances of $R=10\,\Omega$ or smaller,
the value of $V_{0}$ to achieve this is around 2.3~V if the circuit
is biased in the first PDC region and around 3~V is the circuit is
biased in the second PDC region. For a higher resistance such as $R=60\,\Omega$,
these values have to be increased to around 3.95~V and 4.15~V, respectively.
The bias in the second PDC region is convenient for the purpose of
lower power consumption as the steady state current intensity is smaller
in this case. It is important, however, to keep in mind the region
of bistability where both the stable fixed point and limit cycle coexist.
The broader this region is, the more likely bursts of pulses are to
occur when the circuit is perturbed, given the hysteretic character
of the AH bifurcation. Figure \ref{fig:bias_range} shows the width
of the bistability regions in the proximity of both PDC regions versus
the resistance, for two values of $\mu$. For $\mu=0.002\,\Omega^{-1}$,
the bistability has a range in the order of 0.001 V at the left side
and 0.01 V at the right side. For $\mu=0.0002\,\Omega^{-1}$, these
widths decrease in two orders of magnitude. In that sense, biasing
the circuit in the first PDC region is more convenient. The resistance
does not affect the width of the bistability ranges in a substantial
way; only for $R>10\,\Omega$ an increase is observed, although within
the same order of magnitude.

\begin{figure}[th]
\centering{}\includegraphics[width=0.9\columnwidth]{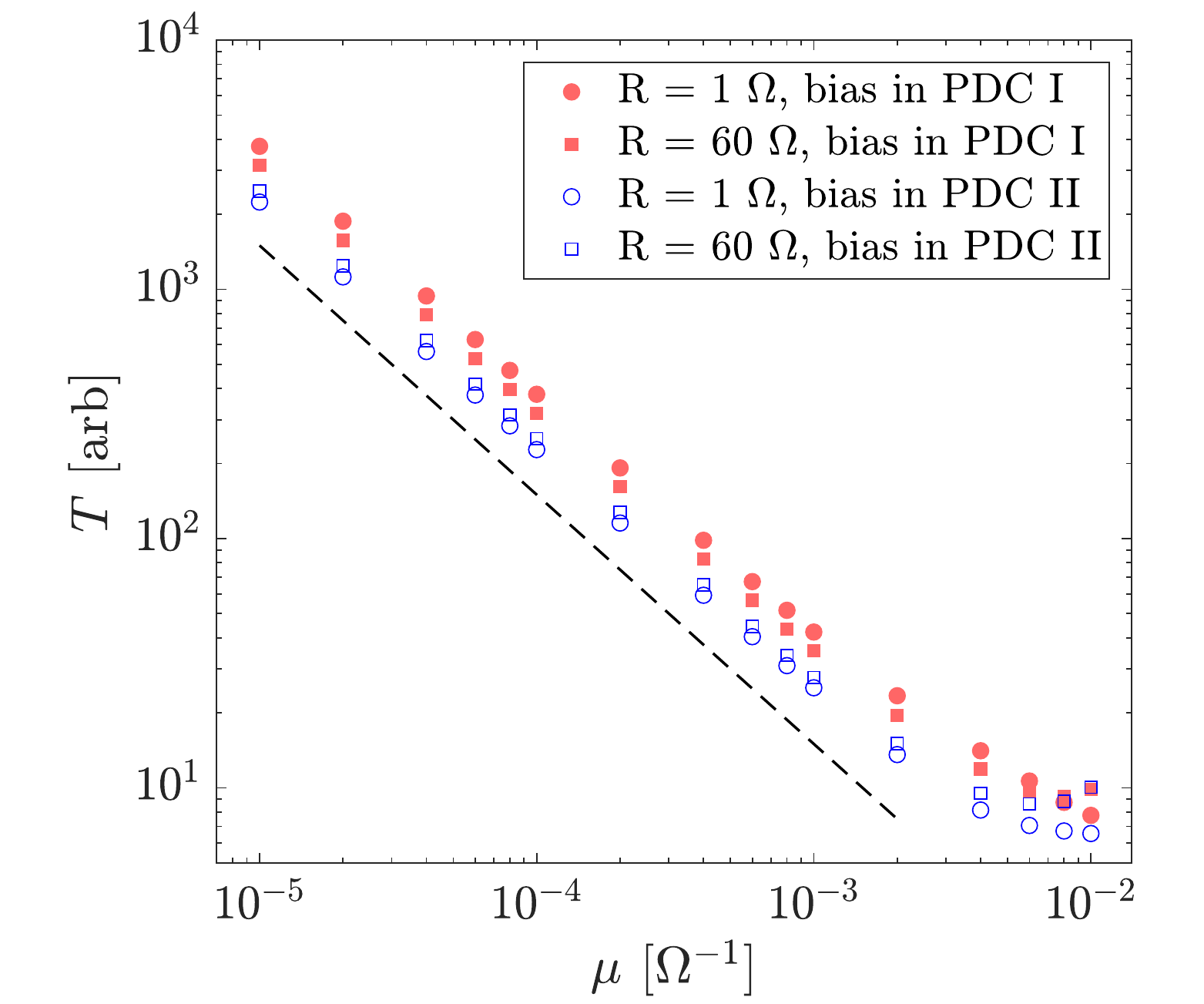}\caption{Period of the stable limit cycle versus the stiffness coefficient
$\mu$, with the circuit biased in the first and second PDC (filled
red and empty blue marks, respectively) for resistances $R$ = 1 $\Omega$
(circles) and $R$ = 60 $\Omega$ (squares). The dashed line has slope
-1 for reference.\label{fig:T_vs_mu}}
\end{figure}

It is also of interest to understand how the lethargic time of the
excitable spikes with the circuit biased at both the first and second
PDC regions is affected by $\mu$ and $R$. An estimation can be provided
by computing the period of the stable limit cycle at the value of
$V_{0}$ corresponding to each AH bifurcation. These calculations
are summarized in figure \ref{fig:T_vs_mu}. Indeed, for $\mu\le0.002\,\Omega^{-1}$,
the period $T$ is observed to decrease like $1/\mu$. This is consistent
with the numerical estimation provided by Romeira et al \citep{RJI-OE-13},
according to which the period --in actual time units-- of stiff,
periodic solutions with slow and fast stages is directly proportional
to the intrinsic inductance. This in turn implies that the period
in normalized time units is inversely proportional to $\mu$. Biasing
the circuit in the second PDC permits spikes with shorter lethargic
times. The resistance has almost no affect on the period; only for
$R$ = 60 $\Omega$ a change is observed (the period decreases in
the first PDC and increases in the second PDC, but all in all, it
remains in the same order of magnitude for the same value of $\mu$).
for $\mu>0.002\,\Omega^{-1}$, the decreasing rate of the period relaxes
and even reverses in the particular case of $R$ = 60 $\Omega$.

\section{Summary and conclusions\label{sec:conclusions}}

A Liénard-type nonlinear oscillator was proposed to model the dynamics
of a double barrier quantum well resonant tunneling diode (DBQW RTD)
connected to an electrical DC input.The configurations where the circuit
behaves as an excitable spike generator were disclosed in a perspective
to design and fabricate optoelectronic, nanoscale devices able of
transmission, reception and storage of spike-coded information. The
RTD oscillator may exhibit one or more equilibrium solutions in the
form of a fixed point or a limit cycle. In particular, the stiffness
coefficients determines whether or not the system behaves as a smooth
oscillator or a spike generator. Indeed, for a sufficiently small
value of $\mu$, the periodic solutions exhibit stages of slow and
fast dynamics, thus producing an electrical output of periodic spikes.

When $V_{0}$ is tuned in the proximity of the NDC, the equilibrium
solution is an attractor point, but perturbations above a certain
threshold trigger an excitable response where and single spike --a
precursor of the periodic solution-- is generated. An electrical
input in the form of white noise generates random spikes, including
bursts of spikes when the device is biased in the second PDC. These
observations are in good agreement with prior experimental results
\citep{RJI-OE-13,RJF-IEEE-13}. We identified the reason for the bursting
as a bistability range between oscillating behavior and rest state.
Bifurcation diagrams at fixed low resistances show a stable limit
cycle along the NDC region, which folds at both ends, becoming unstable.
Both unstable branches coalesce with the fixed point in a subcritical
Hopf bifurcation, producing a bias range of bistability at each boundary
of the NDC region. A smaller stiffness coefficient implies narrower
bias ranges of bistability but a significantly longer period of the
limit cycle and subsequent longer lethargic times for excitable responses,
thus limiting the speed performance. Biasing the device in the first
PDC ensures a narrower bistability range (about one order of magnitude
less than in the second PDC) and, consequently, bursts of spikes and
hysteretic behavior are less likely. However, this choice has the
cost of a higher equilibrium current intensity and, therefore, a higher
power consumption.

Significant qualitative changes occur when the resistance surpasses
a critical value $R_{C}$, given by the minus reciprocal of the minimal
differential conductance at the NDC region (about 38.5 $\Omega$ in
for the I-V characteristic chosen in this work). Depending on the
input bias voltage, the system may exhibit up to three fixed points,
the middle one always being a saddle. In the bifurcation diagram,
this is translated as the fixed point branch folding in two saddle-node
transitions. The saddle point branch and the unstable ends of the
limit cycle branch may coalesce in a homoclinic bifurcation. The device's
intrinsic resistance has little effect on the shape of the spikes,
the lethargic time and the bias ranges. Nonetheless, a low resistance
is preferable for the purpose of low power consumption and to operate
under a configuration where the theoretical model is reliable. Experimentally
measured I-V characteristics in RTDs are not necessarily smooth curves,
or may have more than one NDC \citep{WAO-IEEE-15}, and a non-vertical
load line may intersect it at several points, which may generate a
more complex variety of multistabilities than those observed in this
work.

\section*{Acknowledgements}

This work was supported in part by the European Commission through
the H2020-FET-OPEN Project \textquotedblleft ChipAI\textquotedblright{}
under Grant Agreement 82884. We acknowledge fruitful discussions with
A. Teruel, R. Prohens and C. Vich regarding the dynamics of slow-fast
systems. We thank J. Figueiredo, Universidade de Lisboa, for the fruitful
discussion about RTD devices.


\begin{thebibliography}{10}

\bibitem{Izhikevich}
Eugene~M. Izhikevich.
\newblock {\em Dynamical systems in neuroscience : the geometry of excitability
  and bursting}.
\newblock Computational neuroscience. MIT Press, Cambridge, Mass., London,
  2007.

\bibitem{HH-JOP-52}
A.~L. Hodgkin and A.~F. Huxley.
\newblock A quantitative description of membrane current and its application to
  conduction and excitation in nerve.
\newblock {\em Journal of Physiology}, 117(4):500--544, 1952.

\bibitem{HH2-JOP-52}
A.~L. Hodgkin, A.~F. Huxley, and B.~Katz.
\newblock Measurement of current-voltage relations in the membrane of the giant
  axon of loligo.
\newblock {\em The Journal of physiology}, 116(4):424, 1952.

\bibitem{kuhnert89}
L.~Kuhnert, K.~I. Agladze, and V.~I. Krinsky.
\newblock Image processing using light-sensitive chemical waves.
\newblock {\em Nature}, 337(6204):244--247, 1989.

\bibitem{SNJ-JAP-11}
AS~Samardak, A~Nogaret, NB~Janson, A~Balanov, I~Farrer, and DA~Ritchie.
\newblock Spiking computation and stochastic amplification in a neuron-like
  semiconductor microstructure.
\newblock {\em Journal of Applied Physics}, 109, 2011.

\bibitem{GHR-PRL-07}
D.~Goulding, S.~P. Hegarty, O.~Rasskazov, S.~Melnik, M.~Hartnett, G.~Greene,
  J.~G. McInerney, D.~Rachinskii, and G.~Huyet.
\newblock Excitability in a quantum dot semiconductor laser with optical
  injection.
\newblock {\em Physical Review Letters}, 98:153903, 2007.

\bibitem{SBB-PRL-14}
F.~Selmi, R.~Braive, G.~Beaudoin, I.~Sagnes, R.~Kuszelewicz, and S.~Barbay.
\newblock Relative refractory period in an excitable semiconductor laser.
\newblock {\em Phys. Rev. Lett.}, 112:183902, May 2014.

\bibitem{BKY-OL-11}
Sylvain Barbay, Robert Kuszelewicz, and Alejandro~M. Yacomotti.
\newblock Excitability in a semiconductor laser with saturable absorber.
\newblock {\em Optics letters}, 36(23):4476--4478, 2011.

\bibitem{MAA-SCI-14}
Paul~A. Merolla, John~V. Arthur, Rodrigo Alvarez-Icaza, Andrew~S. Cassidy, Jun
  Sawada, Filipp Akopyan, Bryan~L. Jackson, Nabil Imam, Chen Guo, Yutaka
  Nakamura, Bernard Brezzo, Ivan Vo, Steven~K. Esser, Rathinakumar Appuswamy,
  Brian Taba, Arnon Amir, Myron~D. Flickner, William~P. Risk, Rajit Manohar,
  and Dharmendra~S. Modha.
\newblock {A million spiking-neuron integrated circuit with a scalable
  communication network and interface}.
\newblock {\em {Science}}, {345}({6197}):{668--673}, {Aug 8} {2014}.

\bibitem{Intel-SEEIM-16}
Intel {Q}uark {SE} {M}icrocontroller.
\newblock {\em Intel-SEEIM-4-2016}.

\bibitem{ZZYH-APR-20}
J.~{Zhu}, T.~{Zhang}, Y.~{Yang}, and R.~{Huang}.
\newblock A comprehensive review on emerging artificial neuromorphic devices.
\newblock {\em Applied Physics Reviews}, 7(1):11312, 2020.

\bibitem{ISA-IEEE-17}
R.~{Izumi}, S.~{Suzuki}, and M.~{Asada}.
\newblock 1.98 thz resonant-tunneling-diode oscillator with reduced conduction
  loss by thick antenna electrode.
\newblock In {\em 2017 42nd International Conference on Infrared, Millimeter,
  and Terahertz Waves (IRMMW-THz)}, pages 1--2, Aug 2017.

\bibitem{RFJ-CHAOS-18}
Bruno Romeira, José M.~L. Figueiredo, and Julien Javaloyes.
\newblock Delay dynamics of neuromorphic optoelectronic nanoscale resonators:
  Perspectives and applications.
\newblock {\em Chaos: An Interdisciplinary Journal of Nonlinear Science},
  27(11):114323, 2017.

\bibitem{WAO-IEEE-15}
J.~{Wang}, K.~{Alharbi}, A.~{Ofiare}, H.~{Zhou}, A.~{Khalid}, D.~{Cumming}, and
  E.~{Wasige}.
\newblock High performance resonant tunneling diode oscillators for thz
  applications.
\newblock In {\em 2015 IEEE Compound Semiconductor Integrated Circuit Symposium
  (CSICS)}, pages 1--4, Oct 2015.

\bibitem{OHH-IEEE-16}
N.~{Oshima}, K.~{Hashimoto}, D.~{Horikawa}, S.~{Suzuki}, and M.~{Asada}.
\newblock Wireless data transmission of 30 gbps at a 500-ghz range using
  resonant-tunneling-diode terahertz oscillator.
\newblock In {\em 2016 IEEE MTT-S International Microwave Symposium (IMS)},
  pages 1--4, May 2016.

\bibitem{OHS-IEEE-16}
N.~{Oshima}, K.~{Hashimoto}, S.~{Suzuki}, and M.~{Asada}.
\newblock Wireless data transmission of 34 gbit/s at a 500-ghz range using
  resonant-tunnelling-diode terahertz oscillator.
\newblock {\em Electronics Letters}, 52(22):1897--1898, 2016.

\bibitem{DNN-IEEE-17}
S.~{Diebold}, K.~{Nishio}, Y.~{Nishida}, J.~. {Kim}, K.~{Tsuruda}, T.~{Mukai},
  M.~{Fujita}, and T.~{Nagatsuma}.
\newblock High-speed error-free wireless data transmission using a terahertz
  resonant tunnelling diode transmitter and receiver.
\newblock {\em Electronics Letters}, 52(24):1999--2001, 2016.

\bibitem{WAW-IEEE-18}
Jue Wang, Abdullah Al-Khalidi, Liquan Wang, Razvan Morariu, Afesomeh Ofiare,
  and Edward Wasige.
\newblock 15-gb/s 50-cm wireless link using a high-power compact iii-v 84-ghz
  transmitter.
\newblock {\em IEEE Transactions on Microwave Theory and Techniques},
  66:4698--4705, 2018.

\bibitem{HC-CTA-01}
M.~{Hänggi} and L.~O. {Chua}.
\newblock Cellular neural networks based on resonant tunnelling diodes.
\newblock {\em International Journal of Circuit Theory and Applications},
  29(5):487--504, 2001.

\bibitem{RJF-IEEE-13}
B.~{Romeira}, J.~{Javaloyes}, J.~M.~L. {Figueiredo}, C.~N. {Ironside}, H.~I.
  {Cantu}, and A.~E. {Kelly}.
\newblock Delayed feedback dynamics of liénard-type resonant
  tunneling-photo-detector optoelectronic oscillators.
\newblock {\em IEEE Journal of Quantum Electronics}, 49(1):31--42, Jan 2013.

\bibitem{RJI-OE-13}
B.~Romeira, J.~Javaloyes, C.~N. Ironside, J.~M.~L. Figueiredo, S.~Balle, and
  O.~Piro.
\newblock Excitability and optical pulse generation in semiconductor lasers
  driven by resonant tunneling diode photo-detectors.
\newblock {\em Opt. Express}, 12(8):20931--20940, 2013.

\bibitem{RAF-SR-16}
B.~Romeira, R.~Av{\'o}, Jos{\'e} M.~L. Figueiredo, S.~Barland, and
  J.~Javaloyes.
\newblock Regenerative memory in time-delayed neuromorphic photonic resonators.
\newblock {\em Scientific Reports}, 6:19510 EP --, Jan 2016.
\newblock Article.

\bibitem{LGA-Nature-15}
Yu-Chuan Lin, Ram~Krishna Ghosh, Rafik Addou, Ning Lu, Sarah~M. Eichfeld, Hui
  Zhu, Ming-Yang Li, Xin Peng, Moon~J. Kim, Lain-Jong Li, Robert~M. Wallace,
  Suman Datta, and Joshua~A. Robinson.
\newblock Atomically thin resonant tunnel diodes built from synthetic van der
  waals heterostructures.
\newblock {\em Nature Communications}, 6(1):7311, Jun 2015.

\bibitem{MTC-Nature-14}
A.~Mishchenko, J.~S. Tu, Y.~Cao, R.~V. Gorbachev, J.~R. Wallbank, M.~T.
  Greenaway, V.~E. Morozov, S.~V. Morozov, M.~J. Zhu, S.~L. Wong, F.~Withers,
  C.~R. Woods, Y.-J. Kim, K.~Watanabe, T.~Taniguchi, E.~E. Vdovin,
  O.~Makarovsky, T.~M. Fromhold, V.~I. Fal'ko, A.~K. Geim, L.~Eaves, and K.~S.
  Novoselov.
\newblock Twist-controlled resonant tunnelling in graphene/boron
  nitride/graphene heterostructures.
\newblock {\em Nature Nanotechnology}, 9(10):808--813, Oct 2014.

\bibitem{SDC-EDL-96}
J.~N. {Schulman}, H.~J. {De Los Santos}, and D.~H. {Chow}.
\newblock Physics-based rtd current-voltage equation.
\newblock {\em IEEE Electron Device Letters}, 17(5):220--222, May 1996.

\bibitem{Strogatz}
Steven~H. Strogatz.
\newblock {\em Nonlinear Dynamics and Chaos (with applications to Physics,
  Biology, Chemistry and Engineering)}.
\newblock CRC Press, 2015.

\bibitem{L-RGE-28}
Alfred-Marie Liénard.
\newblock Etude des oscillations entretenues.
\newblock {\em Revue générale de l'électricité}, 23(1):901--912 and
  946--954, 1928.

\bibitem{LGN-PR-04}
B.~Lindner, J.~Garca-Ojalvo, A.~Neiman, and L.~Schimansky-Geier.
\newblock Effects of noise in excitable systems.
\newblock {\em Physics Reports}, 392(6):321 -- 424, 2004.

\bibitem{DDEBT}
K.~Engelborghs, T.~Luzyanina, and D.~Roose.
\newblock Numerical bifurcation analysis of delay differential equations using
  dde-biftool.
\newblock {\em ACM Trans. Math. Softw.}, 28(1):1--21, March 2002.

\bibitem{UTM-IEICE-99}
Tetsuya Uemura and P.~Mazumder.
\newblock Design and analysis of resonant-tunneling-diode (rtd) based high
  performance memory system.
\newblock {\em IEICE Trans. Electron}, E82-C(9), 01 1999.

\end{thebibliography}

\end{document}